\documentclass[twocolumn,nofootinbib,amsmath,prd,aps,superscriptaddress,tightenlines,preprintnumbers]{revtex4}
\usepackage{graphicx}
\usepackage{epstopdf}
\usepackage{amsfonts}
\usepackage{bm}
\usepackage{epsfig}
\usepackage{graphics}
\usepackage{xspace}
\usepackage[usenames]{color}
\def\nn{\nonumber}

\begin{document}

\newcount\hour \newcount\minute
\hour=\time \divide \hour by 60
\minute=\time
\count99=\hour \multiply \count99 by -60 \advance \minute by \count99
\newcommand{\mydate}{\ \today \ - \number\hour :00}

\preprint{UCSD/PTH 08-04}

\title{Electroweak Baryogenesis with a Pseudo-Goldstone Higgs.}

\author{Benjam\'\i{}n Grinstein}
\email[]{bgrinstein@physics.ucsd.edu}
\affiliation{Department of Physics, University of California at San Diego, La Jolla, CA 92093}

\author{Michael Trott}
\email[]{mrtrott@physics.ucsd.edu}
\affiliation{Department of Physics, University of California at San Diego, La Jolla, CA 92093}
\date{\today}                                           

\begin{abstract}
We examine the nature of electroweak Baryogenesis when the Higgs boson's properties are modified 
by the effects of new physics. We utilize the effective potential to one loop (ring improving the finite temperature 
perturbative expansion) while retaining parametrically enhanced dimension six operators
of $\mathcal{O}(v^2/f^2)$ in the Higgs sector. These parametrically enhanced operators 
would be present if the Higgs is a pseudo-goldstone boson of a new physics sector with a 
characteristic mass scale $\Lambda \sim {\rm TeV}$, a coupling constant $4 \, \pi \ge g \ge 1$ and a strong decay 
constant scale $f = \Lambda/g$.  We find that generically the effect of new physics of this form 
allows a sufficiently first order electro-weak phase transition so that the
produced Baryon number can avoid washing out, and has enhanced effects due to new sources CP violation.
We also improve the description of the electroweak phase transition in perturbation theory by determining the  
thermal mass eigenstate basis of the standard model gauge boson fields. This improves the calculation of the finite temperature effects through incorporating mixing in the determination of the vector boson thermal masses of the standard model. 
These effects are essential to determining the nature of the phase transition in the standard model and are of interest in our Pseudo-Goldstone Baryogenesis scenario.

\end{abstract}

\maketitle
\newpage
\section{Introduction}

The Standard Model (SM) of electroweak interactions with a single
Higgs field responsible for spontaneous symmetry breaking is fully
compatible with precision data (EWPD). The theory predicts
the existence of a new particle, the Higgs boson, which has not been
observed. However, the current bound on its mass,   $m_h>114.4 $~GeV
\cite{Barate:2003sz}, is not in conflict with precision tests.

While this is the case, the issues of the hierarchy problem and triviality problems
of the Higgs sector of the SM still strongly motivate
theorists to think that New Physics (NP) will be discovered by LHC at the TeV
scale. EWPD, flavour constraints and the desire for a lack of fine
tuning generically pushes the scale of possible NP degrees of freedom
to $\sim {\rm TeV}$.  EWPD also favors a light Higgs
$m_h \lesssim v$. If the Higgs is part of a NP sector that addresses
the hierarchy and triviality problems with a mass scale $\Lambda \gg v
\gtrsim m_h$, then the model class where the Higgs is a pseudo-goldstone
boson of this NP sector
\cite{Georgi:1975tz,Kaplan:1983sm,Kaplan:1983fs} is an interesting
possibility to consider.\footnote{The Higgs is generally an exact
  goldstone boson in the new sector and receives its mass and self
  couplings from SM corrections in interesting models of this
  form. For a recent review of the physics of pseudo-goldstone Higgs
  models see \cite{Georgi:2007zz}.}

Pseudo-Goldstone Higgs (PGH)
models would be more compelling if the PGH scenario were to address another
problem of the SM $\it{not \, by \, explicit \, construction}$ but as a natural consequence of the 
structure of theory. The purpose of this paper is
to critically examine recent claims that PGH models
can naturally accommodate a very desirable low energy effect, namely the generation of the observed baryon-antibaryon asymmetry of the universe at the electroweak phase transition  (EWPT).

For $\Lambda \gg  v$ it is appropriate to examine the effect of these PGH models on the EWPT using an effective field theory.
This is  appropriate if the scale $\Lambda$ and the details of the new sector are such that all of the NP effects can be 
described by local operators modifying the SM. We are interested in parameter choices of the various models
where this is the case.  We perform a general effective field theory analysis in this paper and are not
wedded to any particular PGH model. Some modern examples of models of this form are little Higgs models \cite{ArkaniHamed:2001nc, ArkaniHamed:2002pa, ArkaniHamed:2002qy} possibly including a custodial symmetry \cite{Chang:2003un}; and Holographic composite Higgs models \cite{Contino:2003ve}
possibly including a custodial symmetry \cite{Agashe:2004rs,Contino:2006qr}. 
Generally speaking, when one imposes custodial symmetry ($\rm SU_C(2)$)
the models can be in accordance with EWPD with a
relatively low $\Lambda$ scale; $\Lambda \sim {\rm TeV}$.

For this reason, the effective theory we use when investigating the low energy effect of PGH models is the SM supplemented with $\rm SU_C(2)$  invariant dimension six operators that 
are $\rm SU(3)\times SU(2) \times U(1)$ invariant and built out of SM fields. In this
paper, we will use Buchmuller and Wyler's \cite{Buchmuller:1985jz} version of this higher dimension operator basis,
although we note that it is over-complete \cite{Mantry:2007sj}. This approach allows us to calculate 
in a relatively model independent manner the lower energy effects of this entire model class.

An important point emphasized in \cite{Giudice:2007fh} when considering dimension six operators induced from a PGH model is that operators that only involve the PGH and derivatives are suppressed by
the decay constant $f$, not the scale $\Lambda$. These scales are related by 
$f = \Lambda/g$ where $g_{sm} \le g \le 4 \, \pi$ is the coupling constant of the new sector.  This parametric enhancement of the effects of NP in the Higgs sector (when $g>1$) make the properties of the Higgs a very important 
probe of such models.

This parametric enhancement is important when considering electroweak Baryogenesis (EWB) as many of the constraints on EWB are constraints on the self couplings of the Higgs. In the SM, these directly translate into a constraint on the Higgs mass. In \cite{Grinstein:2007iv} the significant effect of NP (such as PGH models) on the 
relationship between the Higgs self coupling and the Higgs mass was studied in detail. This turns out to be crucial in the Pseudo-Goldstone Baryogenesis (PGBG) scenario we examine in this paper, as it will allow the phase transition to be sufficiently first order
for the produced baryon-antibaryon asymmetry to avoid washing out while the Higgs mass is greater than 114.4 $\rm GeV$. If the PGH model also contains new sources of CP violation, these sources of CP violation are also parametrically enhanced and PGBG could occur for a scale $f$ in the range $500 \, {\rm GeV} \lesssim f \lesssim 1 \, {\rm TeV}$. 

 Our results agree with some aspects of a recent study by Wells {\it et\,al} \cite{Delaunay:2007wb} although we do find some disagreements and improve upon the analysis in a number of ways (see Section \ref{past}). The most important improvements are that we determine and use a thermal mass eigenstate basis of the gauge boson fields of the SM that distinguishes between the transverse and longitudinal masses; and we also use a canonically normalized Lagrangian incorporating the effects of the parametrically enhanced 
kinetic sector operators. This latter improvement expands the allowed parameter space for PGBG considerably.

The outline of this paper is as follows. In Section \ref{sectwo} we review the deficiencies of the SM with regards to EWB. In Section \ref{higherd} we state the effective theory that we will use to investigate the low energy
effects of PGH models.  In Section \ref{models} 
we review some PGH models that can induce the Wilson coefficients of interest.
In Section \ref{effP} we derive the effective potential including finite temperature effects and improve the 
SM calculation of finite temperature masses.  
In Section \ref{phase} we examine the EWPT in the low energy effective theory and determine when it can be first order. 
In Section \ref{washout} we discuss the necessary conditions on the low energy realization of PGH models so that the 
produced baryon asymmetry will not wash out and briefly discuss the low energy effect on bubble nucleation.
In Section \ref{conc}  we conclude.

\section{SM Baryogenesis at the EW scale}\label{sectwo}

Following the initial realization that anomalous EW baryon number violation is large at high temperatures \cite{Linde:1977mm,Dimopoulos:1978kv}, the possibility of EWB \cite{Kuzmin:1985mm} proceeding through a first order EWPT was suggested.
This suggestion grew into a promising theory \cite{Cohen:1993nk}, 
see \cite{Quiros:1999jp} and references therein for a 
summary. However, this elegant theoretical
mechanism is now ruled out for the SM with one
Higgs doublet.
A brief review of the problems of EWB in the SM is appropriate before we reexamine EWB in the context of NP.
In the SM, Baryogenesis at the EW scale has a number of serious problems.  
For Baryogenesis to take place, the Sakharov conditions \cite{Sakharov:1967dj}
\begin{itemize}
\item [(i)] C and CP violation,
\item [(ii)]  baryon number violation,
\item [(iii)] a departure from thermal equilibrium,
\end{itemize}
must be satisfied. All three of these conditions are qualitatively present in the SM with a light Higgs.
C \cite{tHooft:1976up,tHooft:1976fv} and CP violation are present \cite{Kobayashi:1973fv}. A departure from thermal equilibrium and 
baryon number violation was potentially present \cite{Kuzmin:1985mm} in the EWPT if the Higgs self coupling was sufficiently small. In the SM this requires a small Higgs mass, $m_h \lesssim 70 \,{\rm GeV}$.

In part due to the Higgs mass bound $m_h > 114.4 \, {\rm GeV}$ \cite{Barate:2003sz}, 
it is known that the SM
alone, quantitatively, does not generate the correct  baryon asymmetry
observed in the universe. The current value of the baryon asymmetry at
$95 $ \% CL  is determined from BBN cosmology \cite{PDBook} 
and three year WMAP data \cite{Spergel:2006hy}, to be
\begin{eqnarray}
Y_B = \frac{\rho_B}{s} = 
\begin{array}{cc}
(6.7,9.2) \times 10^{-11} & {\rm BBN} , \\
(8.1,9.2) \times 10^{-11} & {\rm WMAP}, 
\end{array}
\end{eqnarray}
where $\rho_B$ is the Baryon number density of the universe and $s$ is the entropy density of the universe.
The number to compare for the SM  baryon asymmetry due to EWB is zero. 
This is due to the following problems.
\begin{itemize}
\item [(P1)] The EWPT  in the SM is not first order as the Higgs mass exceeds 70 {\rm GeV}. 
This statement is based on lattice simulations \cite{Kajantie:1996qd,Kajantie:1996mn,Kajantie:1995kf,Csikor:1998eu}
and similar conclusions are reached in perturbative studies. Therefore, the required departure from thermal equilibrium is not present in the SM.
\item [(P2)]  Even if the EWPT was weakly first order, EWB could not occur. The  EWPT must be strongly first order
for the resulting $Y_B$ to not wash out and this requires a Higgs mass that is quite small.
Washout can occur as thermal Boltzmann fluctuations can erase a generated $Y_B$. These fluctuations depend on the energy of the EW vacuum barrier field configurations \cite{Carson:1990jm,Shaposhnikov:1987tw} as $\exp{(-E_{sph}/T_c)}$ (where $E_{sph}$ is the sphaleron energy) and translate into the following bounds on the expectation value of Higgs field at the critical temperature of the phase transition $T_c$ 
\begin{eqnarray}
\frac{\langle \phi (T_c) \rangle}{T_c} \gtrsim b.
\end{eqnarray}
Here $b$ is a numerical constant estimated to be in the range $1.0\lesssim
b\lesssim 1.3$ \cite{Dine:1991ck} from the uncertainty in the
caluclation of the functional determinant  associated with the static saddle point solution of the Yang-Mills Higgs equations, the sphaleron.
This solution translates into a constraint on the Higgs mass in the SM as \cite{Dine:1992vs}
\begin{eqnarray}
\frac{\langle \phi (T_c) \rangle}{T_c} \sim \frac{4 \, E \, v^2}{m_h^2},
\end{eqnarray}
where $E = (4 \,  m_w^3 + 2 \, m_z^3)/(12 \, \pi \, v^3)$ which gives
\begin{eqnarray}
m_h \lesssim v \, \sqrt{\frac{4 \, E}{b}}, 
\end{eqnarray}
yielding bounds of $m_h \lesssim 35 \, {\rm GeV}$ for $b = 1.3$ and $m_h \lesssim 39 \, {\rm GeV}$ for $b = 1.0$. \footnote{The 
more optimistic end of this estimated bound  $\phi (T_c)/{T_c} \gtrsim 1$ is frequently used in the literature. We will treat the constraint as exact and consider both ends of the bound when we consider the effects of NP.}

\item [(P3)] Finally, CP violation in the SM is far too small for EWB \cite{Gavela:1993ts,Gavela:1994dt,Huet:1994jb}. 
\end{itemize}

\subsection{New mechanism or Higgs effective field theory?}

Comparing these problems with our baryon-antibaryon asymmetric existence, one could conclude 
that a totally new mechanism such as leptogenesis \cite{Fukugita:1986hr} or perhaps supersymmetry \cite{Carena:1996wj}
is involved in the generation of $Y_B$. Leptogenesis is also an appealing mechanism and will become much more so if 
neutrinoless double beta decay \cite{KlapdorKleingrothaus:2001ke} is unambiguously established in future experiments.
The window for the MSSM to allow EWB is now quite constrained \cite{Quiros:2000wk}. MSSM Baryogenesis requires a Higgs mass $m_h \lesssim 115 \, {\rm GeV}$.

Another solution to problems P1, P2 could be that the Higgs would have properties that deviate from the SM.  If the Higgs couples to a NP sector with a mass scale $\Lambda \sim {\rm TeV}$ then the properties of the Higgs are naturally expected to deviate from the SM below the scale $\Lambda$. The question in detail becomes to what degree must the properties of the Higgs effective field theory (HEFT) Higgs deviate from the SM Higgs for EWB  to occur, and how natural is the required deviation in a NP setting?

The naive expectation that the properties of the effective theory Higgs will deviate insignificantly from the SM Higgs fails in many model extensions of the SM, see \cite{Mantry:2007ar,Mantry:2007sj,Randall:2007as,Grinstein:2007iv, Noble:2007kk, Manohar:2006gz, O'Connell:2006wi,Fan:2008jk,Goldberger:2007zk,Profumo:2007wc}. In particular, in PGH models with a new strong interaction at a $\rm TeV$, there is a parametric enhancement of the NP effects on the properties of the Higgs \cite{Giudice:2007fh}. When this is the case, the properties of the Higgs in the effective theory can change dramatically and the problems of EWB can be addressed as follows.

\begin{itemize}
 \item [(P1)] The relationship between the Higgs mass and the Higgs self coupling is significantly relaxed in the effective theory \cite{Grinstein:2007iv}. A first order EWPT is possible due to a small effective Higgs self coupling while $m_h>114.4 $~GeV; we will show this by determining the required conditions on the dimension six operator Wilson coefficients so that the EWPT is first order. To accomplish this we utilize ring improved finite temperature  perturbation theory and analytically study the EWPT in our effective theory.
We also determine the relevant thermal mass eigenstates for the 
scalar and gauge boson masses that are crucial to determining the nature of the phase transition. This improves the SM 
calculation of the EWPT and is of interest in our HEFT.

 \item [(P2)] We also determine the constraint on the coefficients of dimension six operators from the washout condition in our effective theory. There are several effects that modify the determined value of $\langle \phi(T_c) \rangle$ in the effective theory.
We include Higgs self energy loops which are large and neglected in the expression $E = (4 \,  m_w^3 + 2 \, m_z^3)/(12 \, \pi \, v^3)$.
Constraints on the Higgs mass are significantly effected as NP changes the order of the polynomial of the effective potential, and the thermal mass basis we derive significantly modifies the determined value.  \item [(P3)] We are assuming that the Higgs is a pseudo-goldstone boson of a new sector which can have the required new sources of CP violation. The low energy expression
of the new CP violation is through operators that are also parametrically enhanced. The
required scale suppressing the NP operators for the SM to be supplemented with enough CP violation is in the range $500 \, {\rm GeV} \lesssim f
\lesssim 1 \, {\rm TeV}$ according to recent independent studies \cite{Pospelov:2005pr,Huber:2006ri}. This does not contradict the current bounds 
on non SM CP violation from EDM experiments, see \cite{Pospelov:2005pr,Huber:2006ri}. 
\end{itemize}

One would expect  $500 \, {\rm GeV} \lesssim f \lesssim 1 \, {\rm TeV}$  in PGH models with a new strong interaction at $\sim \rm TeV$. 
When $f$ is in this range, the effective Higgs self coupling can be small enough for a strong EWPT while $m_h> 114.4 \, {\rm GeV}$. With this outline of our approach in mind, we first determine the operator basis that expresses the low energy effects
of a new strong interaction at a $\rm TeV$ in the next section.

\section{The Lagrangian Density with $D= 6$  Higgs Operators  }\label{higherd}

We now construct  the Lagrangian density of a HEFT due to integrating out the degrees of freedom with masses greater than $v$ of a PGH 
model.  The SM Lagrangian density is given by
\begin{eqnarray}
\label{LHiggs}
{\mathcal{L}}_{\phi}^4 =  
\left(D^\mu \,\phi \right)^\dagger \, \left(D_\mu \, \phi \right)  - V \left( \phi \right),
\end{eqnarray}
where $\phi$ is the Higgs scalar doublet.  The covariant derivative of the
$\phi$ field is given by
\begin{eqnarray}
D_{\mu} = 1 \, \partial_\mu - i \, \frac{g_1}{2} \, B_\mu - i \, g_2 \,  \frac{\sigma^I}{2} \, W_\mu^I,
\end{eqnarray}
where $\sigma^I$ are the pauli matrices, $W_\mu^I,B_\mu$, are the $\rm SU(2)$
and $\rm U(1)$ SM gauge bosons and the hypercharge of $1/2$ has been
assigned to the Higgs.  The Higgs potential at tree level is given by
\begin{eqnarray}
V(\phi) =  -m^2 \, \phi^\dagger \, \phi  + \frac{\lambda_1}{2} \, \left( \phi^\dagger \phi \right)^2.
\end{eqnarray}

We expand the real field $h(x)$ around a real constant background field value $\varphi$ in Landau gauge introducing three real Goldstone boson fields $\chi_i(x)$
\begin{eqnarray}
\phi(x) = \frac{1}{\sqrt{2}} \,
\left(
 \begin{array}{c} 
 \chi_1(x) + i \, \chi_2(x) \\
\varphi+ h(x) + i \, \chi_3(x) 
\end{array}
\right).
\end{eqnarray}
When the tree level masses are fixed by a minimization of the SM potential, they are given by
\begin{eqnarray}
m_h^2(\varphi) &=& \frac{\lambda_1}{2} \left(3 \varphi^2 - v^2 \right), \quad m_\chi^2(\varphi) = \frac{\lambda_1}{2} \left(\varphi^2 - v^2 \right), \nn \\
m_W^2(\varphi) &=& \frac{g_2^2 \, \varphi^2}{4}, \quad m_Z^2(\varphi) = \frac{(g_1^2 + g_2^2) \, \varphi^2}{4}, \\  
m_i^2(\varphi) &=& \frac{f_i^2 \, \varphi^2}{2}. \nn
\end{eqnarray}
In the SM one has $\varphi = v \equiv \sqrt{2 \, m^2/\lambda_1}$.

We utilize Landau gauge as this gauge choice allows us to 
avoid subtleties that occur in unitary gauge
in finite temperature field theory, see \cite{Dolan:1973qd,Sher:1988mj}.
The gauge fixing is performed by taking $\xi \rightarrow 0$ for the Lagrangian term
\begin{eqnarray}
\mathcal{L}_{gauge} &=& - \frac{1}{2 \, \xi}(\partial^\mu \, W_\mu^i - \frac{\xi}{2} \, g_2 \, \varphi \, \chi^i)^2 \nn \\
&\,& -  \frac{1}{2 \, \xi}(\partial^\mu \, B_\mu - \frac{\xi}{2} \, g_1 \, \varphi \,(h + i \, \chi_3))^2.
\end{eqnarray}

We now turn to the low energy effect of PGH models inducing parametrically enhanced higher dimension operators.  The effective Lagrangian density with operators that contain Higgs doublets is 
\begin{eqnarray}
\mathcal{L}_{\phi} = {\mathcal{L}}_{\phi}^4 + \frac{{\mathcal{L}}_{\phi}^6}{{\Lambda}^2} +  {\mathcal{O}}(\frac{v^4}{{\Lambda}^4}),
\end{eqnarray}
with the dimension six Lagrangian density  (recall $g$ is the coupling constant of the new sector and $ g_{sm} \le g \le 4 \, \pi$)
\begin{eqnarray} \label{custodialeqn}
{\mathcal{L}}_{\phi}^6 &=& g^2 \,  C_{\phi} \, \partial^\mu \, ( \phi^\dagger \, \phi) \partial_\mu \, ( \phi^\dagger \, \phi)
-  g^2 \, \frac{\lambda_2}{ 3 \, ! } \, \left( \phi^\dagger \, \phi \right)^3  \\
&\,& + \frac{C_{h G}}{2} (\phi^\dagger \, \phi) \, G_{\mu \, \nu} \, G^{\mu \, \nu} 
 + \frac{C_{h \tilde{G}}}{2} (\phi^\dagger \, \phi) \, G_{\mu \, \nu} \, \tilde{G}^{\mu \, \nu} \nn \\
&\,& + \frac{C_{h W}}{2} (\phi^\dagger \, \phi) \, W_{\mu \, \nu} \, W^{\mu \, \nu} 
+ \frac{C_{h \tilde{W}}}{2} (\phi^\dagger \, \phi) \, W_{\mu \, \nu} \, \tilde{W}^{\mu \, \nu} \nn \\
&\,& + \frac{C_{h B}}{2} (\phi^\dagger \, \phi) \, B_{\mu \, \nu} \, B^{\mu \, \nu} 
+ \frac{C_{h \tilde{B}}}{2} (\phi^\dagger \, \phi) \, B_{\mu \, \nu} \, \tilde{B}^{\mu \, \nu} + \cdots \nn
 \end{eqnarray}
Where the Wilson coefficients are independent of $g$.
We have written the custodial symmetry \cite{Susskind:1978ms,Sikivie:1980hm}  ($\rm SU(2)_C$) preserving
terms involving only Higgs doublets and field strengths. As mentioned, approximate custodial symmetry is favored 
as it will suppress the T parameter operator $(\phi^\dagger \, D^{\mu} \, \phi)^2$ that contributes to the $\rho$ parameter  \cite{Giudice:2007fh, Grinstein:2007iv}. The PDG quotes $\rho_0 = 1.0002^{+ 0.0007}_{-0.0004}$ for
the global fit \cite{PDBook} of EWPD and this operator is only 
suppressed by the decay constant scale.  The coefficient of this operator has been determined \cite{Barbieri:2004qk,Grinstein:2007iv} to be $C < 4 \times 10^{-3}$ for $f = 1 \, {\rm TeV}$.
\footnote{The need for a  $\rm  SU_C(2)$ symmetry in models of this form (and many other SM extensions) has been appreciated for quite some time and many PGH models, such as \cite{Georgi:1984af,Chang:2003un,Contino:2006qr} incorporate this symmetry by construction.}

Thus if $\rm  SU_C(2)$ is 
not approximately preserved in extensions of the SM of the form we are discussing, this operator would have to be  
suppressed by fine tuning or the decay constant scale would have to be quite high. With an approximately $\rm  SU_C(2)$  invariant NP sector however, the 
strong decay constant could be as low as $f = \Lambda/g \sim 500 \, {\rm GeV}$. In this case, the effects on the 
Higgs sector self couplings are very significant \cite{Grinstein:2007iv}. This is the scenario we are interested in. We do not consider this to be a strong assumption as  $\rm  SU_C(2)$ is also approximately preserved in the SM.

In the SM, $\rm SU_C(2)$ is only an approximate symmetry as custodial symmetry is violated by the $\rm U(1)$ and Yukawa interactions. 
The dimension six operators involving modifications of the Yukawa sector of the SM 
provide a further source of CP violation required for EWB, see \cite{Pospelov:2005pr,Huber:2006ri}. These operators and their hermitian conjugates are also suppressed by $f$ and are given by
\begin{eqnarray}
O_{e \, \phi} &=& g^2 (\phi^\dagger \, \phi) ( \bar{\ell} \, e \, \phi), \\
O_{u \, \phi} &=& g^2 (\phi^\dagger \, \phi) ( \bar{q} \, u \, \phi), \\ 
O_{d \, \phi} &=& g^2 (\phi^\dagger \, \phi) ( \bar{q} \, d \, \phi). 
\end{eqnarray}
So long as $f$  is in the range
$500 \, {\rm GeV} \lesssim f \lesssim 1 \, {\rm TeV}$ these operators can supply the extra CP violation for EW Baryogenesis in our PGBG scenario. 

Further distinctions can be made on the ${\mathcal{L}}_{\phi}^6$ operator basis. All of the operators of interest can come from underlying tree level topologies  \cite{Arzt:1994gp}, thus their Wilson coefficients need not be suppressed by factors of $16 \, \pi^2$ in NDA \cite{Manohar:1983md}. One expects the field strength operators that must be induced by loops to be significantly suppressed compared to the parametrically enhanced operators in the Higgs sector for this reason.\footnote{For PGH theories, phenomenological
signals involving field strength operators have been examined in \cite{Giudice:2007fh}.}

We also note that an important aspect of PGH models that has been neglected in some of the 
literature on the EWPT \cite{Bodeker:2004ws,Delaunay:2007wb,Grojean:2004xa,Noble:2007kk} is that operator extensions of the SM induce a non-canonical effective Lagrangian. This is  due to the presence of dimension six kinetic operators.
As in \cite{Barger:2003rs,Grinstein:2007iv,Giudice:2007fh} we normalize the kinetic term of the resulting Lagrangian for $h$ to $1/2$. We use the field redefinition 
\begin{eqnarray}
 h(x)  \to \frac{h'(x)}{(1+ 2 \, \frac{\varphi^2}{f^2} \,C_{\phi})^{1/2}}.  
\end{eqnarray} 
This gives the potential, before minimization fixes $\varphi$, the form 
\begin{eqnarray}
V_{C}(h',\chi_i')  &=& \nn \\
&\,& \hspace{-2cm} - \frac{m^2}{2} \,\left( \sum _{i = 1,3}\chi_i^2 +  \left[\varphi + \frac{h'}{(1+ 2 \, \frac{\varphi^2}{f^2} \,C_{\phi})^{1/2}} \right]^2 \right) \nn \\
&\,& \hspace{-2cm}  + \frac{\lambda_1}{8}  \,\left( \sum _{i = 1,3} \chi_i^2 +  \left[\varphi + \frac{h'}{(1+ 2 \, \frac{\varphi^2}{f^2} \,C_{\phi})^{1/2}} \right]^2 \right)^2  \\
&\,& \hspace{-2cm} + \frac{\lambda_2}{48 \,f^2} \,\left( \sum _{i = 1,3} \chi_i^2 +  \left[\varphi + \frac{h'}{(1+ 2 \, \frac{\varphi^2}{f^2} \,C_{\phi})^{1/2}} \right]^2 \right)^3 \nn
\end{eqnarray} 
Neglecting constant terms and expanding in the $f \rightarrow \infty$ limit while retaining only 
$1/f^2$ terms one has for $h$
\begin{eqnarray}
V_{C}(h')  &=& a_h(\varphi) \, (\varphi^3) \,  h' +\frac{m_h^2(\varphi)}{2} h'^2  + \frac{\varphi \, \lambda_3^{eff}}{3 \, !} \, h'^3 \nn \\
&+&  \frac{\lambda_4^{eff}}{4 \, !} \, h'^4  + \frac{15 \, \lambda_2}{5 \, !  \, f^2} \, \varphi \,  h'^5 + \frac{15 \, \lambda_2}{6 \, ! \, f^2} \,  h'^6, 
\end{eqnarray} 
where the parameters $a_h(\varphi),m_h^2(\varphi)$ and the effective couplings are given by
\begin{eqnarray}
\label{effmass1}
a_h(\varphi) &=& \frac{\lambda_1}{2} \left(1 - \frac{v^2}{\varphi^2} \right) \left(1 - \frac{\varphi^2}{f^2} \,C_{\phi} \right) + \frac{\lambda_2}{8 \, f^2} \left(\varphi^2 - \frac{v^4}{\varphi^2} \right) \nn \\
m_h^2(\varphi) &=&  \frac{\lambda_1}{2}\left(3  \varphi^2 - v^2\right) \left(1 - 2  \frac{\varphi^2}{f^2} \,C_{\phi} \right) + \frac{\lambda_2}{8 \, f^2} \left(5 \, \varphi^4 - v^4 \right) \nn  \\
\label{lambda3eff1}
\lambda_3^{eff}(\varphi)&=& 3 \, \lambda_1\, \left(1 - 3 \, \frac{\varphi^2}{f^2} \,C_{\phi} \right)  + \frac{5}{2} \, \lambda_2 \, \frac{\varphi^2}{f^2}\nn \\
\label{lambda4eff1}
\lambda_4^{eff}(\varphi)&=& 3 \, \lambda_1 \, \left(1 - 4 \, \frac{\varphi^2}{f^2} \,C_{\phi} \right)  +  \frac{15}{2} \, \lambda_2 \, \frac{\varphi^2}{f^2}.\nn 
\end{eqnarray} 
Note that we have eliminated $m$ with Eqn (\ref{npminima}).  We use the zero temperature minimization condition that determines $m^2$ even though we are interested in the inclusion of finite temperature effects. The inclusion of finite temperature effects shifts the mass for the Higgs field but does not change the minimization condition that determines $m^2$ as the UV subtraction is defined when $T =0$. Also, we neglect the higher order Coleman-Weinberg terms in the minimization condition. Similarly for each $\chi_i$ field one has
\begin{eqnarray}
V_{C}(\chi_i') &=&\frac{m_\chi^2(\varphi)}{2} \, (\chi_i')^2  + 
(\chi_i')^4 \left[ \frac{\lambda_1}{8} + \frac{\lambda_2 \, \varphi^2}{16 \, f^2} \right] 
+  \frac{(\chi_i')^6  \, \lambda_2}{48 \, f^2} \nn
\end{eqnarray} 
where we have
\begin{eqnarray}
m_\chi^2(\varphi) = \frac{ \lambda_1}{2} \left(\varphi^2 - v^2\right) + \frac{\lambda_2}{8 \, f^2} \left( \varphi^4 - v^4 \right), 
\end{eqnarray} 
and there are many cross terms in the potential that we have not written here for brevity.

We suppress the primes on the redefined fields for the remainder of the paper. 
For the background field $\varphi$ we have
\begin{eqnarray}
V_{C}(\varphi) &=&  - \frac{m^2}{2} \varphi^2 + \frac{\lambda_1}{8} \, \varphi^4 + \frac{\lambda_2}{48 \, f^2} \, \varphi^6.
\end{eqnarray}

Once we obtain the full effective potential in Section \ref{phase}, it will be clear that neglecting to reduce the effective Lagrangian to a canonical form will effect the $\langle \varphi \rangle$. 
Canonically normalizing the effective Lagrangian also effects the
crucial relationship between the Higgs mass and the Higgs self couplings. As both the effects of $C_{\phi}$ and $\lambda_2$ are enhanced by the same parameter ($g^2$), one should not neglect the effects of
canonically normalizing the HEFT Lagrangian when examining PGBG.

With this effective theory,
we can investigate the low energy effects of PGH models on the EWPT and the washout condition.  Firstly, we examine some examples of integrating out degrees of freedom with mass scales $\Lambda > v$ and matching to induce the 
HEFT we are discussing
\section{Model Particulars} \label{models}

As we have explained above, our analysis is based on a Lagrangian  with
precisely the same field content as that of the SM. It has been
supplemented with additional terms, irrelevant operators chracterized
by the dimensionfull paramter $f$. The advantage of this approach is
that one can study the conditions for succesful Baryogenesis without
specifying a specific ``ultaviolet completion,'' that is, without
committing to one specific model of interactions beyond the standard
model. All that is required is then that the model includes a light
higgs and that the parameters of the resulting low energy effective
Lagrangian fall in a certain range, as will be shown in Fig. 7-10.

It is easy to display simple UV completions of the effective
Lagrangian under study. A minimalistic example consists of the SM
supplemented by a neutral, real scalar field $S$ and additional terms
in the Lagrangian
\begin{equation}
\Delta{\cal L}=\frac12\partial^\mu S\partial_\mu S-\frac12 f^2S^2-\frac1{3!}\kappa_1fS^3-\kappa_2fS\phi^\dagger\phi.
\end{equation}
A quartic self-interaction for $S$ can be added to make the potential
bounded from bellow but is irrelevant for our purposes.  Integrating out the scalar $S$ at energies below its mass $f$
one obtains an effective potential of the form of Eq.~(\ref{custodialeqn}) with the couplings
\begin{eqnarray}
\lambda_1 &=&  \lambda-\frac{\kappa_2^2}{2}\\
\frac{g^2 \lambda_2}{\Lambda^2} & =& -\frac{\kappa_1\kappa_2^3}{f^2}\\
\frac{g^2C_{\phi}}{\Lambda^2} & =&\frac{\kappa_2^2}{2 \, f^2}, 
\end{eqnarray}
where $\lambda$ is the quartic higgs self-coupling before the effects
of the $S$ field are included. Hence we see that this simple model
produces, at lowest order, only the terms in the effective Lagrangian
that play a significant role in our analysis of Baryogenesis but does
not give any other terms including notably those which could be
significantly constrained by precision tests of the EW sector.
This model of course does not address the hierarchy problem. 
Models that involve a new strong interaction can address the hierarchy
problem and can also generate the EW scale through dimensional 
transmutation and are more compelling.

The possibility that a light higgs is a composite particle whose
constituents are bound by a new interaction that goes strong at a
scale $\sim1$~TeV has been extensively studied; for a review see \cite{Georgi:2007zz}.
In most of these models the higgs mass remains small compared to the
scale of the new strong interactions because it is a pseudo-Goldstone
boson of a global symmetry broken only weakly (typically by the EW and
yukawa interactions in the SM). We can determine which of these models
have succesful EW Baryogenesis by determining their effective
Lagrangian. In particular, we need to know the magnitude of $f^2$,
$\lambda_2$ and $C_{\phi}$. As we shall see, models in which the precision EW constraints are
evaded by adopting a large scale, $f\sim10$~TeV, require unacceptably
large coefficients of  $\lambda_2$ and $C_{\phi}$. However, since
the precision EW constraints are most severe for the $\rho$ (or $T$)
parameter, the scale $f$ can be taken significantly smaller in models
with an $SU_C(2)$ symmetry that automatically supresses
corrections to $\rho$. Let us consider some examples of strongly coupled pseudo-goldstone higgs theories with custodial symmetry.

The Littlest Higgs with custodial symetry \cite{Chang:2003zn} is a theory with flavor
symmetry $SO(9)$ in which it is assumed that techni-strong
interactions induce a condensate that breaks flavor to $SO(5)\times
SO(4)$. An $SU(2)^3\times U(1)$ subgroup of the flavor group is gauged
weakly, but some of these gauged symmetries are spontaneously broken at
the scale of the condensates so that, in fact, only the SM gauge group
remains unbroken and, of the original goldstone bosons, only the higgs doublet remains light. 
At low energies this model is of the type we are investigating, with 
the couplings
\begin{eqnarray}
\frac{g^2 \lambda_2}{\Lambda^2} & \simeq& \frac{\lambda_1}{f^2}, \\
\frac{g^2C_{\phi}}{\Lambda^2} & =&\frac1{4\, f^2}.
\end{eqnarray}
In the above the exact expression for $\lambda_2$ has not been computed; the expression above satisfies the counting rules of \cite{Giudice:2007fh}.

The Holographc Higgs model \cite{Contino:2003ve,Agashe:2004rs,Contino:2006qr}
is a warped 5D theory with 4D-boundaries. A bulk
$SO(5)\times U(1)\times SU(3)$ gauge symmetry is broken  to
$O(4)\times U(1)\times SU(3)$ on the UV boundary and to the SM on the
IR one. Matching to the low energy 4D effective theory
gives\cite{Giudice:2007fh}
\begin{eqnarray}
\frac{g^2 \lambda_2}{\Lambda^2} & =& c\frac{\lambda_1}{f^2} \\
\frac{g^2 C_\phi}{\Lambda^2} & =&\frac1{2\, f^2}
\end{eqnarray}
where $c=0,1$ in the
models of Refs.~\cite{Agashe:2004rs} and~\cite{Contino:2006qr}  respectively.
\footnote{Models of this form can also supply a dark matter candidate \cite{DiazCruz:2007be} and can also increase the strength of the
phase transition through other 5D effects in Gauge-Higgs unification \cite{Maru:2006wx}.}

In these examples of matching we find that, firstly, the symmetry breaking that 
induces $\lambda_2$ is proportional to $\lambda_1$; and  secondly, $C_\phi$ is positive.
However, our small number of examples are in no way exhaustive of all PGH models.
In particular, models of the little higgs form, by construction can have
a symmetry breaking structure that is quite surprising due to collective symmetry breaking, see \cite{Georgi:2007zz}. Thus we will 
perform our effective theory analysis in two cases.
\begin{itemize}
\item [C1]: In the first case we will retain the maximum model 
independence that can be afforded in the PGH structure and allow $\lambda_2$ to be independent
of $\lambda_1$.
\item [C2]:  In the second case we will impose that $\lambda_2$ is proportional to $\lambda_1$
and determine a constraint on $\tilde{\lambda}_2 = \lambda_2/\lambda_1$ and $C_{\phi}$.
\end{itemize}

We now turn to the construction of our effective potential for the low energy effective theory of PGH models.

\section{Effective Potential} \label{effP}

We now calculate the effective potential to one loop
to determine the leading quantum corrections to the classical tree level potential.
Studies of this form were inaugurated by Coleman and Weinberg in 
\cite{Coleman:1973jx} and several reviews of the application of the effective potential 
in studies of the electroweak phase transition exist \cite{Sher:1988mj,Quiros:1999jp}. 
As well as the one loop temperature independent terms, there are also one loop finite temperature terms 
determined using thermal field theory, see \cite{kapusta:2006pm,Kajantie:1995dw,Rubakov:1996vz}.  
First we consider the temperature independent effective potential.

\subsection{One Loop Effective Potential}

The effective potential is determined as the sum of 1PI diagrams with arbitrary numbers of external legs and zero external momenta as shown in Fig.\ 1. We will renormalize the one loop contributions to the effective potential 
term by term using dim reg with $ d= 4- 2 \, \epsilon$ and $\overline{\rm MS}$.
\begin{figure}[bh] \label{effpot}
\centerline{\scalebox{0.8}{\includegraphics{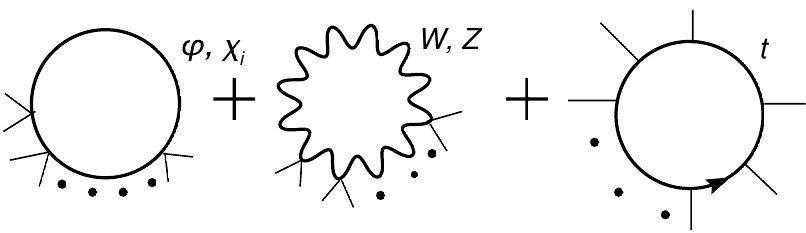}}}
\caption{One loop diagrams that contribute to the effective potential.}
\end{figure}

We neglect terms due to higher dimension operator insertions in the above loop diagrams when determining the zero temperature effective potential. In the next section and in the Appendices we 
do include the effects of NP in thermal loops. We do this as the latter are significantly numerically enhanced and 
have an important thermal screening effect on the one loop effective potential. This reduces the problem with the imaginary part of the effective potential as we will show.  These NP effects in thermal loops and the effects of NP that change the relationship between the Higgs mass and the self couplings in the HEFT are the dominant effects of NP that we are investigating.

\subsection{Scalar Contributions}
The effective potential is determined in terms of the classical background field $\varphi$. 
The contributions of the Higgs self interactions to the one loop effective potential are given by
\begin{eqnarray}
V^{eff}_{S,h}(\varphi) =  \frac{\mu^{4 - d}}{2} \, \int \frac{d^4 k_E}{(2 \pi)^4} \, \ln \left(k_E^2 + V''_{C}(0) \right),
\end{eqnarray} 
where we have introduced the renormalization scale $\mu$. Note that we have rotated to Euclidean space. 
From the previous section we have $ V''_{C}(0) = m_h^2(\varphi)$. 
We perform the integral to obtain, 
\begin{eqnarray}
V^{eff}_{S,h}(\varphi) &=&  - \frac{m_h^4(\varphi)}{32 \, \pi^2} \, \frac{4 \, \Gamma(2 - d/2)}{d(d-2)} \, \left(\frac{m_h^2(\varphi)}{4 \, \pi \, \mu^2} \right)^{d/2-2},\nn
\end{eqnarray} 
where $V''_{C}$ indicates two derivatives with respect to the dynamical field $h$. We find
\begin{eqnarray}
V^{eff}_{S,h}(\varphi) &=& \frac{m_h^4(\varphi)}{64 \, \pi^2} \, \left[\log \left(\frac{m_h^2(\varphi)}{\mu^2}\right)  - \frac{3}{2}  - C_{uv}  \right],\nn
\end{eqnarray} 
where
\begin{eqnarray}
C_{uv} = \frac{1}{\epsilon} - \gamma_E + \log(4 \, \pi).
\end{eqnarray} 
There are also contributions from the three $\chi_i$ fields for the scalar contribution to the effective potential.
Each $\chi_i$ field gives a contribution 
\begin{eqnarray}
V^{eff}_{S,\chi_i}(\varphi) &=& \frac{m_\chi^4(\varphi)}{64 \, \pi^2} \, \left[\log \left(\frac{m_\chi^2(\varphi)}{\mu^2}\right)  - \frac{3}{2}  - C_{uv} \right]. \nn
\end{eqnarray} 

\subsection{Vector Bosons and Fermions}
The one loop effects due to the spinors that receive their
mass from the vacuum expectation value of the Higgs are well known, see \cite{Quiros:1999jp} for a review. The one loop results are given by  
\begin{eqnarray}
V^{eff}_{F}(\varphi) &=& - \sum_i \, \frac{3 \, m_i^4(\varphi)}{16 \, \pi^2} \, \left[\log \left(\frac{m_i^2(\varphi)}{\mu^2}\right) - \frac32  - C_{uv} \right] \nn
\end{eqnarray} 
We neglect all but the top quark contributions. Due to the operator $O_{t \, \phi}$ there are also $1/f^2$ corrections of the form 
$\sim {\rm Re}[C_{t \, \phi}] \,\varphi^2/f^2$ to the mass, see \cite{Mantry:2007sj} for a recent study of these operator effects.
As these contributions to the potential are suppressed by $\varphi^2/( 16 \, \pi^2 \,f^2)$ we neglect them.
For the $W$ and $Z$ fields one obtains
\begin{eqnarray}
V^{eff}_{V}(\varphi) 
&=& \frac{ 3 \, m_Z^4(\varphi)}{64 \, \pi^2} \, \left[ \log\left( \frac{m_Z^2(\varphi)}{\mu^2} \right)
 -  \frac56 - C_{uv}  \right], \nn \\
&+& \frac{3 \, m_W^4(\varphi)}{32 \, \pi^2} \, \left[ \log\left( \frac{m_W^2(\varphi)}{\mu^2}  \right)
 -  \frac56 - C_{uv}  \right] \nn
\end{eqnarray} 

The one loop contribution to the effective potential for our low energy theory is thus
\begin{eqnarray}
V^{eff}(\varphi) &=& V_{C}(\varphi) + V_{S,h}^{eff}(\varphi) + 3 \,  V_{S,\chi_1}^{eff}(\varphi) \nn \\
&\,&  + V^{eff}_{V}(\varphi) +V^{eff}_{F}(\varphi). \nn
\end{eqnarray} 

\subsection{Renormalization}

We are using $\overline{\rm MS}$ and dimensional regularization to define the UV subtraction
in the $T \rightarrow 0$ limit. 
The UV counter terms are given by
\begin{eqnarray}
\mathcal{L}_{c.t.} = \delta \, \Omega + \delta \, m^2 \, \varphi^2 + \delta \, \lambda_1 \, \varphi^4
\end{eqnarray}
where the counterterm parameters are given by
\begin{eqnarray}
\delta \, \Omega &=& \frac{m^4}{16 \, \pi^2} \,  C_{uv} \nn \\
\delta \, m^2 &=& - \frac{3 \, \lambda_1 \, m^2}{64 \, \pi^2} \,  C_{uv}  \\
\delta \, \lambda_1 &=&  \frac{3 \, C_{uv}}{64\, \pi^2}  \, \left[ \lambda_1^2  \, 
 -  f_t^4 + \frac{g_2^4}{8} + \frac{(g_1^2 + g_2^2)^2}{16} \right] \nn
\end{eqnarray}
The first renormalization condition defines the vacuum expectation value of $\varphi$. 
Although we will retain an unfixed $\varphi$ when examining the electroweak phase transition, 
as a check of our results so far we can minimize the potential in the $T \rightarrow 0$ limit while neglecting the 
higher order effects of the one loop effective potential terms.
We interpret the effective potential as a function of $\Phi = h + \varphi$ and minimize with respect to $\Phi$
and take $ \langle \chi_i \rangle = 0, \langle h  \rangle=0$ and $\langle \Phi \rangle = v$.
Solving for $m^2$ in the minimization condition up to neglected $\mathcal{O}(\varphi^4/f^{4}, \varphi^2/(f^2 \, 16 \, \pi^2))$ 
terms in the one loop effective potential one finds
\begin{eqnarray}\label{min}
\frac{m^2}{\varphi^2} = \frac{\lambda_1}{2} + \frac{\lambda_2 \, \varphi^2}{8 \, f^2}
\end{eqnarray} 
The solution of this equation for the classical minimum is 
\begin{eqnarray}\label{npminima}
\varphi^2 = v^2 \equiv \frac{2 f^2}{\lambda_2} \, \left[- \lambda_1 \pm \sqrt{\lambda_1^2 + 2 \, \lambda_2 m^2/f^2} \right]
\end{eqnarray}

With this definition, we find for the $h$ field 
\begin{eqnarray}
V_{C}(h)  &=& \frac{m_h^2}{2} \, h^2  + \frac{v \, \lambda_3^{eff}}{3 \, !} \, h^3 + \frac{\lambda_4^{eff}}{4 \, !} \, h^4 + \frac{15 \, \lambda_2}{5 \, !  \,f^2} \, v \,  h^5 \nn \\
&\,& + \frac{15 \, \lambda_2}{6 \, ! \, f^2} \,  h^6
\end{eqnarray} 
where the parameters are
\begin{eqnarray}\label{higgsmass}
\frac{m_h^2(v)}{v^2} &=& \lambda_1\, \left(1 - 2 \, C_{\phi} \, \frac{v^2}{f^2} \right)+
\frac{\lambda_2}{2} \, \frac{v^2}{f^2},  \\
\lambda_3^{eff}(v)&=& 3 \, \lambda_1\, \left(1 - 3 \,  C_{\phi} \, \frac{v^2}{f^2} \right) 
 + \frac{5}{2} \, \lambda_2 \, \frac{v^2}{f^2}, \\
\lambda_4^{eff}(v)&=& 3 \, \lambda_1 \, \left(1 - 4 \, C_{\phi} \, \frac{v^2}{f^2} \right)  + \frac{15}{2} \, \lambda_2 \, \frac{v^2}{f^2}.  
\end{eqnarray} 

For the $\chi_i$ fields, the minimized potential causes the mass of the fields to vanish, as expected.

In Appendix $\rm A$ we derive a range of values for $|\lambda_1|$. The largest values that $|\lambda_1|$ can take on while the loop expansion is under control are $|\lambda_1| \sim g_2^3$. We now examine one aspect of how this power counting
affects the computation of the effective potential. The effective potential and hence its derivatives
correspond to Green functions with vanishing external momenta ($P^2=0$). Conversely, the physical parameters are defined at the 
scale $m_h^2$. Thus formally we have
\begin{eqnarray}
 \frac{d^2 V^{eff}}{d \, \Phi^2} &=& \hat{m}_h^2 - \Delta \, \Sigma \nn \\
 \frac{d^3 V^{eff}}{d \, \Phi^3} &=& \hat{\lambda}^{eff}_3 - \Delta \, \Gamma_3 \nn \\
 \frac{d^4 V^{eff}}{d \, \Phi^4} &=& \hat{\lambda}^{eff}_4 - \Delta \, \Gamma_4 \nn 
\end{eqnarray}
where we are following and extending the convention laid down in \cite{Delaunay:2007wb} and hats denote physical parameters, for example,
$\hat{m}_h$ is the pole of the Higgs propagator. 
We have introduced the shifts for the 1PI $2,3$ and $4$ point functions
\begin{eqnarray}
\Delta \, \Sigma &=& \Sigma(P^2 = m_h^2) - \Sigma(P^2 = 0), \nn \\
\Delta \, \Gamma_3 &=& \Gamma_3(P^2 = m_h^2) - \Gamma_3(P^2 = 0), \nn \\
\Delta \, \Gamma_4 &=&  \Gamma_4(P^2 = m_h^2) - \Gamma_4(P^2 = 0).
\end{eqnarray}
to denote this discrete running of the parameters. 

Note that if $|\lambda_1| \lesssim g_{2}^3$  holds we should neglect the small effects due to this shift in the 
parameters as 
\begin{eqnarray}
\Sigma(P^2 = m_h^2) - \Sigma(P^2 = 0) &\sim& g_{1,2}^2 \, p^2, \nn \\
&\sim& g_{1,2}^2 \,m_h^2, \\
&\sim& g_{1,2}^2 \, v^2 \left( \lambda_1 + \frac{ \lambda_2 \, v^2}{2f^2} \right). \nn \\
\end{eqnarray}
and similarly 
\begin{eqnarray}
\Gamma_3(P^2 = m_h^2) - \Gamma_3(P^2 = 0) &\sim&  g_{1,2}^2 \, v \,  \lambda_3^{eff}, \\
\Gamma_4(P^2 = m_h^2) - \Gamma_4(P^2 = 0) &\sim& g_{1,2}^2 \, \lambda_4^{eff}.
\end{eqnarray}
Clearly if $|\lambda_1| \lesssim g_{2}^3$ we can neglect $g_{1,2}^2 \, \lambda_1$ effects. We also must neglect $ g_{1,2}^2 \, \lambda_2$ effects as these
are loop suppressed and suppressed by $f^2$; we have neglected many such effects in the effective potential
and  consistency demands 
that we drop these terms.  If one chooses to retain these terms because $\lambda_1$ is not small then these effects can be significant. However, at the same time the convergence of the loop expansion will be poor and perturbative investigations will be limited in the reliability of their conclusions as shown in Appendix $\rm A$.

As advocated in \cite{Anderson:1991zb,Delaunay:2007wb} it can be important to determine the running of the parameters in the Higgs sector to 
formally cancel the IR divergence that occurs when $\varphi \rightarrow v, T \rightarrow 0$. The correct description of the $T \rightarrow 0$ physics of the system after the EW phase transition is complete should 
cancel this IR divergence. However,  as this is not our focus in this paper we neglect this higher order effect and renormalize in the standard manner using $\overline{\rm MS}$ and dimensional regularization. 

A much more significant effect when $|\lambda_1| \lesssim  g_{2}^3$ is distinguishing between the transverse and longituginal masses of the gauge bosons and determming the thermal mass basis for the SM appropriate for ring resummation. This is an  $\mathcal{O} (g_{1,2}^2 \,  m \,  T)$ effect that imposes significant physical constraints when extensions to the SM still have $\lambda_1, m^2 >0$
and is numerically important in our HEFT. We now turn to finite temperature effects and determining the 
transverse and longituginal thermal mass basis in the SM.

\subsection{Finite Temperature Effects}\label{finiteT}

The finite temperature effects are calculated using fields with
(anti)periodic boundary conditions for the (fermion)boson fields
on the time interval $\beta = 1/T$ \cite{fradkin}. These boundary conditions allow one to  decompose the 
Bose ($\varPhi$) and Fermion ($\varPsi$) fields in fourier modes \cite{Rubakov:1996vz}
\begin{eqnarray}
\varPhi(x,\tau) &=& \sum_{n = -\infty}^{\infty} \phi_n(x) \, \exp(i \, \omega^B_n \, \tau), \nn \\
\varPsi(x,\tau) &=& \sum_{n = -\infty}^{\infty} \psi_n(x) \, \exp(i \, \omega^F_n \, \tau), \nn 
\end{eqnarray}
where we have $\omega^B_n = 2 \,  n \, \pi \, T$ and $\omega^F_n = (2 \, n + 1)  \pi \, T$. 
The one loop functions $J_{\pm}$ are obtained  \cite{Dolan:1973qd,Weinberg:1974hy} 
by using residues to transform the 
sum over fourier modes into the sum of the usual $T=0$ loop contributions to propagators 
(which are renormalized in the standard way) and additionally finite temperature contributions 
that have correction factors for the Fermi- Dirac and Bose-Einstein particle distributions. The temperature dependent contributions are written in terms of the integrals \cite{Dolan:1973qd}
\begin{eqnarray}
J_{\pm}(y_i^2) \equiv \int_0^\infty d x \, x^2 \, \log \left[{1 \mp \, \exp{\left(- \sqrt{x^2 + y_i^2} \right)}} \right],
\end{eqnarray}
where $y^2_i = m^2_i/T^2$. The temperature dependent one loop terms are given by
\begin{eqnarray} \label{finitet}
V_{T}(\varphi,T) = \frac{T^4}{2 \, \pi^2} \left(- \sum_{F} \, g_f  \, J_{-}(y^2_i) + 
\sum_{B} \, g_B  \, J_{+}(y^2_i) \right),
\end{eqnarray} 
where the sums are taken over all degrees of Boson (B) and Fermion (F) freedom. The number of degrees of freedom $g_i$ for the $W^{\pm},Z,t, h, \chi_i$ fields are
\begin{eqnarray}
g_t &=& 12, \quad g_{W^{\pm}} = 6, \quad g_Z = 3, \\
g_h &=& 1, \quad g_{\chi} = 3. \nn
\end{eqnarray}

The Fermi- Dirac and Bose-Einstein particle distribution correction factors modify the loop expansion parameter.
The finite temperature loop expansion is an expansion in
\begin{eqnarray}
\frac{g_{SM}^2}{\exp^{(- E \, \beta)} \pm 1},
\end{eqnarray}
where $E$ is the typical energy scale of a process and one has a $+(-)$ sign for Fermi-Dirac(Bose-Einstein) particle distributions \cite{Gross:1980br} .
As $T \gg E$ the effective expansion parameter  for the Fermions is still given by 
$g_{SM}^2$. However, for the bosonic degrees of freedom, as $T \gg E$, the expansion parameter is given by
\begin{eqnarray}
\frac{g_{SM}^2 \,T}{E}.
\end{eqnarray}
Thus at high temperature, perturbation theory begins to break down in the Bosonic loops.
This fact is essential to the phenomena of high 
temperature symmetry restoration.
Otherwise, perturbative corrections (for all $T$) would never restore EW
gauge symmetry at high temperatures.  

The IR divergence $T >> E$ driven breakdown
of finite temperature field theory is decidedly inconvenient in perturbative studies of the EW phase transition. A mathematical sign of this breakdown is the presence of a $m^3(\varphi) \, T$ term in the high temperature expansion of the finite temperature integral for the bosons. 
We resum a class of higher order diagrams that act to introduce a thermal mass $\propto T^2$ which screens the   
IR divergence in the bosonic propagators \cite{Carrington:1991hz}. This ring resummation improves the nature of the thermal perturbative expansion and can be formally justified by a power counting analysis \cite{Quiros:1999jp,Weinberg:1974hy,Fendley:1987ef,Espinosa:1993yi} which considers the tadpole diagrams calculated at finite temperature shown in Fig. 2. 
A scalar tadpole, at leading order in $M/T$  gives a finite temperature contribution 
\begin{eqnarray}
\lambda_1 \frac{T^2}{4}.
\end{eqnarray}
Consider adding $n$ quadratically divergent subdiagrams to a tadpole as in Fig. 2.
This diagram will scale as 
\begin{eqnarray}
\left( \frac{\lambda_1 \, T^2}{M^2} \right)^n \lambda_1 T \, M = (\lambda_1)^{2} \, \frac{T^{3}}{M} \, \left(\frac{\lambda_1 T^2}{M^2} \right)^{n-1}.
\end{eqnarray}
\begin{figure} 
\centerline{\scalebox{1.2}{\includegraphics{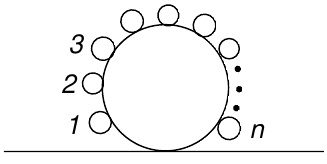}}}
\caption{The $n$ tadpole loop contribution to the diasy diagram of the Higgs propagator.}
\end{figure}
For temperatures where $\lambda_1 T^2 \sim M^2$ this class of ring diagrams 
should be resummed for a reliable perturbative expansion. In fact, in our HEFT, the temperature scales  of the EWPT are such 
that this factor is typically less than one. However, ring resummation is still an important improvement on the 
naive thermal perturbative expansion as employing ring resummation improves the convergence of the loop 
expansion \cite{Weinberg:1974hy,Arnold:1992rz} and reduces the imaginary part of the effective potential.

The imaginary part of the effective potential is of concern as our description of the phase transition assumes that the field is sufficiently stable for the transition to be described by bubble nucleation. So long as $\rm Im(V_{eff}) \ll Re(V_{eff})$, the imaginary part can be interpreted following \cite{Weinberg:1987vp} as the decay rate per unit volume of a state, see also \cite{Sher:1988mj}. The imaginary part of the effective potential can potentially come from two sources. The logarithms of the Coleman-Weinberg terms when a mass squared is negative and the cubic mass terms that appear in the expansion of the $J_{+}$.  The logarithmic dependence on the mass cancels 
when finite temperature effects are included as the finite temperature
integrals $J_{\pm}$ {\it are} the Coleman-Weinberg terms regulated with a finite temperature cut off \cite{Quiros:1999jp} and ring resummation cures the remaining imaginary part in the following manner.  Consider the Higgs mass 
\begin{eqnarray}
m_h^2(\varphi) = \frac{\lambda_1}{2} \, \left( 3 \, \varphi^2 - v^2 \right),
\end{eqnarray}
when $m^2$ is eliminated. Typically $\varphi \ll v$ until far after the phase transition has occurred and this term is negative before thermal corrections are taken into account.  When performing a ring resummation, we rewrite the potential as the standard one loop finite temperature contributions $V_{T}(\varphi,T)$ and then an extra term that includes the shift in the mass due to the ring resummation thermal corrections, $\Pi_h(T) = \lambda_1 \, T^2/4 + T^2 \, B_T$ 
where
\begin{eqnarray}
B_T= \frac{4 \, f_t^2 + 3 g_2^2 + g_1^2}{16}.
\end{eqnarray}
See Appendix $\rm B$ for details. For the Higgs, we have
\begin{eqnarray}
V_R(\varphi,T) = \frac{T}{4 \, \pi^2} \, \int_0^{\infty} \, dk \, k^2 \log \left[1 + \frac{\Pi_h(T)}{k^2 + m_h^2(\varphi)} \right].
\end{eqnarray}
This integral can be directly evaluated using the Leibniz rule for $m_h^2$, integrating over $k$, and subsequently integrating over $m_h^2$; one finds
\begin{eqnarray}
V_R(\varphi,T) = \frac{T}{12 \, \pi} \left[m_h^3(\varphi) - (m_h^2(\varphi) + \Pi_h(T))^{3/2} \right].
\end{eqnarray} 
The first term cancels against an equivalent cubic mass term in the high temperature expansion of $J_+$ for the Higgs.
The remaining term is crucial in determining the nature of the EWPT. The expression $m_h^2(\varphi) + \Pi_h(T)$ is an example of a thermal eigenstate mass.
Thus we see that ring resummation introduces a thermal mass term so that  
\begin{eqnarray}
m_h^2(\varphi,T) = \frac{\lambda_1}{2} \, \left( 3 \, \varphi^2 - v^2 \right) + \frac{\lambda_1 \, T^2}{4} + T^2 \, B_T,
\end{eqnarray}
for the cubic mass dependence. As we are studying the phase transition as $T$ decreases from values $T \sim v$ these thermal mass terms introduce a
positive real contribution for the mass that makes $m_h^2(\varphi)$ positive for the range of $\varphi,T$ of interest in our study of the effective potential. A similar argument holds for all bosonic degrees of freedom \cite{Carrington:1991hz}.
The class of diagrams suppressed by
single power of $\lambda_1$ compared to the diagrams of the ring resummation are the `setting sun'
diagrams, see Fig.3. These diagrams are not included in
the resummation which is justified {\it so long as $| \lambda_1 |$ is small}.
\begin{figure} 
\centerline{\scalebox{1.2}{\includegraphics{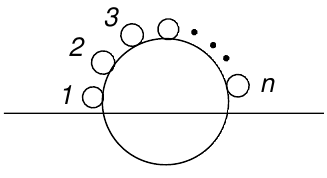}}}
\caption{The $n$ tadpole loop contribution to the sunset diagram of the Higgs propagator.}
\end{figure}

The key difference in this part of our analysis compared to the past literature is taking thermal contributions to the gauge boson polarization tensor self consistently into account and defining a thermal mass eigenstate basis for the gauge boson fields. Past calculations have neglected this subtlety in defining the mass eigenstates in the context of thermal corrections. 
This improvement is of numerical importance as the phase transition occurs when all the terms in the effective potential are approximately the same size and are canceling against one another. The critical value of the vev $\langle \varphi(T_c) \rangle$ and the critical temperature $T_c$  that determine the washout criteria are sensitive to small changes in these mass terms. We discuss at length the calculation of the gauge boson mass eigenstate basis in Appendix $\rm B2$.  
We find the following longitudinal vector boson masses
\begin{eqnarray}
(m^L_W(\varphi,T))^2 &=&  g_2^2 \, \left(\frac{11 \, T^2}{6} + \frac{\varphi^2}{4} \right), \\
(m^L_A(\varphi,T))^2 &=&   \frac{11  T^2}{6}  \left(g_1^2 \,  \cos^2(\theta(T)) + g_2^2\,  \sin^2(\theta(T))  \right) \nn \\
&\,& + \frac{\varphi^2}{4} \left(g_1 \, \cos(\theta(T)) - g_2 \, \sin(\theta(T))\right)^2, \nn \\
(m^L_Z(\varphi,T))^2 &=&  \frac{11 T^2}{6} \left(g_2^2 \,  \cos^2(\theta(T)) + g_1^2\,  \sin^2(\theta(T))  \right) \nn \\
&\,& + \frac{\varphi^2}{4} \left(g_1 \, \sin(\theta(T)) +g_2 \, \cos(\theta(T))\right)^2. \nn
\end{eqnarray}
where we have introduced a thermal Weinberg angle $\theta(T)$ that characterizes the 
degree of mixing in the longitudinal vector boson masses.
For the transverse masses (determined again in Appendix $\rm B2$) we incorporate the effects of mixing 
and introduce a second thermal Weinberg angle $\theta'(T)$ characterizing the 
degree of mixing in the transverse vector boson masses. We also introduce a 
parameter $\gamma$ that signifies a nonperturbative magnetic mass term
that is important as it screens the transverse mass of the $W,Z$ fields in the $\varphi \rightarrow 0$ limit.
We will use the value $\gamma= 4.2$ which has been determined for the
deconfined hot $\rm SU(2)$ gauge theory \cite{Heller:1997nqa} in Landau gauge. We expect this to be a good approximation to the 
$\gamma$ of the SM.
We find the following transverse masses
\begin{eqnarray}
(m^T_W(\varphi,T))^2 &=&\frac{\gamma^2 \, g_2^4}{9 \, \pi^2} \, T^2 + \frac{g_2^2 \, \varphi^2}{4}, \\
(m^T_A(\varphi,T))^2 &=&   \frac{g_1^2 \, T^2 \, \cos^2[\theta'(T)]}{24} \nn \\
&+&  \frac{\varphi^2 \left(g_2 \, \sin[\theta'(T)] - g_1 \, \cos[\theta'(T)] \right)^2}{4}, \nn \\
(m^T_Z(\varphi,T))^2 &=& \frac{g_2^2 \, \, m^T_W(\varphi,T) \, T \, \cos^2[\theta'(T)]}{3 \, \pi} \nn \\
&+&  \frac{g_1^2 \, T^2 \, \sin^2[\theta'(T)]}{24} \nn \\
&+&  \frac{\varphi^2 \left(g_2 \, \cos[\theta'(T)] + g_1 \, \sin[\theta'(T)] \right)^2}{4}. \nn
\end{eqnarray}
Note that both of our thermal Weinberg angles $\theta(T),\theta'(T)$ reduce to $\theta_W$ in the $T \rightarrow 0$ limit.

In our HEFT we also have to deal with the effects of NP on the ring resummation for the Higgs and
the would be goldstone boson fields $\chi_i$. Again examining the 
Higgs mass we have
\begin{eqnarray}
m_h^2(\varphi) =  \frac{\lambda_1}{2}\left(3  \varphi^2 - v^2\right) \left(1 - 2  \frac{\varphi^2}{f^2} \,C_{\phi} \right) + \frac{\lambda_2}{8 \, f^2} \left(5 \, \varphi^4 - v^4 \right). \nn
\end{eqnarray}
Thus NP can make matters worse in a number of ways. If $\lambda_2$ and $\lambda_1$ are independent, the $T^2$ thermal mass term is small compared to the term proportional to $- \lambda_2 \, v^4/f^2$ ($\lambda_1$ is at best $\sim g_2^3$ whereas, when $\lambda_1$ and $\lambda_2$ are independent, we allow $\lambda_2 \sim \mathcal{O}(1)$). So the mass is negative and the effective potential is not dominated by its real part near the EWPT. Further, for large regions of parameter space in PGBG $\lambda_1 < 0$ so that the quadratic thermal corrections make the situation worse. When this occurs one must have $\lambda_2 > 0 $ to stabalize the potential. Both of these problems are solved (and the loop expansion is improved) if one also incorporates the ring diagrams proportional to 
$\lambda_2$. This introduces terms of the form $\lambda_2 \, T^4/f^2$ and  $\lambda_2 \, T^2 \, \varphi^2/f^2$ that act to ensure that the Higgs and would be goldstone boson fields have a positive mass for the $\varphi,T$ of interest. We perform this calculation in Appendix $\rm B1$.

We find the following expressions for the masses appropriate
for the ring resummed effective potential for the $h$ and $\chi_i$
\begin{eqnarray}
m_h^2(\varphi,T) &=& m_h^2(\varphi) + \frac{T^2 \, \lambda_1}{4} \left(1 - 3 \, C_{\phi} ^1 \frac{\varphi^2}{f^2} \right)
+ \frac{\varphi^2 \, T^2}{2 \, f^2} \, \lambda_2 \nn \\
&+& T^2 \, B_T \,   \left(1 - 4 \, C_{\phi} ^1 \frac{\varphi^2}{f^2} \right) + \frac{3 \, T^4 \, \lambda_2}{4 \, f^2}, \nn \\
m_\chi^2(\varphi,T)  &=&m_\chi^2(\varphi) + \frac{T^2 \, \lambda_1}{4} \left(1 -  C_{\phi} ^1 \frac{\varphi^2}{f^2} \right)
+ \frac{\varphi^2 \, T^2}{2 \, f^2} \, \lambda_2 \nn \\
&+& T^2 \, B_T  + \frac{3 \, T^4 \, \lambda_2}{4 \, f^2}. 
\end{eqnarray}
Note that we neglect the $\mathcal{O}(g^3_{SM} \, T)$ and $\mathcal{O}(\lambda_1 \, g_{SM} \, T)$ loop suppressed
contributions from the one loop gap equations \cite{Buchmuller:1993bq} for the scalars.

We note that the ring resummation utilizes the result of the high temperature expansion of the $J_{\pm}(y_i^2)$ and $\Pi_h(T)$ is approximated by its leading $T^2$ term. In our effective theory the critical temperature $T_c$ at which the minima of the potential become degenerate can be significantly less than the EW scale $v$. Further, the effects of supercooling due to the expansion of the universe delaying the onset of the phase transition lead to the physically relevant nucleation temperature $T_n < T_c$  being, in some cases, $T_n \lesssim m_h, m_W, m_Z$  \cite{Delaunay:2007wb}. As this is the case, one might question the general use of the high temperature expansion in this analysis and others. In particular, one might doubt the convergence of the expansion used in Appendix $\rm B$ to determine the thermal mass  basis.

However, this approach is under control\footnote{Once again, this statement holds  {\it so long as $| \lambda_1 |$ is small}. When $|\lambda_1| \gg  0.2$ generally the temperatures $T_n,T_c$ are too small for a reliable high temperature expansion. This is another sign of the lack of a consistent perturbative treatment for large negative $\lambda_1$.}
 for thebosonic fields when we expand the $J_+$ integral as
\begin{eqnarray}\label{expj}
J_{+}(y^2) &=& \frac{\pi^2 \, y^2}{12} - \frac{\pi \, (y^2)^{3/2}}{6} - \frac{y^4}{32} \, \log \left[\frac{y^2}{a_b} \right],
\end{eqnarray}
where $\log a_b = 5.408$, and $\log a_f = 2.635$.
Taking into account the effects of the 
expansion of the universe we are restricted to the situation where $ \varphi_c \sim T_c$ and 
$T_n$ is not too far below $T_c$ for most of the ($\lambda_2, C_{\phi}$) parameter space of interest. Lower temperatures lead to metastable vacuum solutions and the EWPT does not occur, see \cite{Delaunay:2007wb}. 
Using Eqn.~(\ref{expj}) is sufficiently accurate, so long as  $m_i/T < 2 \, \pi$. The 
neglected higher order terms are a numerically suppressed expansion given by
\begin{eqnarray}
- 2 \, \pi^{7/2} \, \sum_{\ell = 1}^\infty \,(-1)^\ell \frac{\zeta(2 \, \ell + 1)}{(\ell + 1)!} \, \Gamma(\ell + 1/2) \, \left(\frac{m^2}{(4 \, \pi^2 \, T^2)} \right)^{\ell +2}.
\end{eqnarray} 
Thus Eqn.~(\ref{expj}) is clearly sufficient for 
all known masses as the lowest physically interesting temperatures are $T \sim 20 \, {\rm GeV} $. For the unknown Higgs mass, we restrict ourselves to considering low Higgs masses $m_h \lesssim160 \, {\rm GeV}$ for this reason.

\subsection{The Effective Potential in the PGH effective theory} \label{phase}
We find the following effective potential
\begin{eqnarray}\label{effpot}
V_{eff}^{ring}(\varphi, T) &=& \frac{a}{2} \left(T^2 - T_b^2 \right) \, \varphi^2 + \frac{\lambda_2}{48 \, f^2} \, \varphi^6
+ \overline{\lambda_1}(T,f) \, \varphi^4, \nn \\ 
&-& \frac{T}{12 \, \pi} \, \left[ m_h^3(\varphi,T) + 3 \, m_\chi^3(\varphi,T) \right], \nn \\
&-&  \frac{T}{12 \, \pi} \, \left[(m_A^L)^3(\varphi,T) + (m_Z^L)^3(\varphi,T)\right], \nn \\
&-&  \frac{T}{12 \, \pi} \, \left[ 2 \, (m_W^L)^3(\varphi,T) + 2 \, (m_A^T)^3(\varphi,T)  \right], \nn \\
&-&  \frac{T}{12 \, \pi} \, \left[2 \, (m_Z^T)^3(\varphi,T) + 4 \, (m_W^T)^3(\varphi,T)\right], \nn \\
&+& L(\varphi,T) + \mathcal{O} \left(g_{sm}^4, \lambda_1^2, f_t^4, \frac{\varphi^2}{16 \, \pi^2 \, f^2}\right).
\end{eqnarray}
We have used the condition
\begin{eqnarray}
\frac{\partial \, V_{eff}(v,0)}{\partial \, \varphi} \equiv 0,
\end{eqnarray}
to fix $m^2$  in defining the potential and have adopted the notation 
\begin{eqnarray}
a &=&  B_T + \frac{\lambda_1}{4} \left(1+ \frac{v^2}{3 \, f^2} \, C_\phi \right), \\  
T_b^2 &=&  m^2/a, \nn \\
&=& \left(\frac{\lambda_1 \, v^2}{2} + \frac{\lambda_2 \, v^4}{8 \, f^2}\right)/a, \\
\overline{\lambda_1}(T,f) &=& \frac{\lambda_1}{8} + \frac{\lambda_2 \, T^2}{24 \, f^2} - \lambda_1 \, C_\phi \, \frac{T^2}{8 \, f^2}, 
\end{eqnarray}
and the logarithmic terms are given by
\begin{eqnarray}
L(\varphi,T) &=& - \frac{3 f_t^2}{64 \, \pi^2} \, \varphi^4 \left({\rm log} \left[\frac{a_f \, T^2}{\mu^2} \right] - \frac32 \right) \nn \\
&\,& + \frac{3 \, m_\chi^4(\varphi,0)}{64 \, \pi^2} \, \left({\rm log} \left[\frac{a_b \, T^2}{\mu^2} \right] - \frac32 \right) \nn \\
&\,& + \frac{m_h^4(\varphi,0)}{64 \, \pi^2} \, \left({\rm log} \left[\frac{a_b \, T^2}{\mu^2} \right] - \frac32 \right) \nn \\
&\,& + \frac{3 \, m_W^4(\varphi,0)}{32 \, \pi^2} \, \left({\rm log} \left[\frac{a_b \, T^2}{\mu^2} \right] - \frac56 \right) \nn \\
&\,& + \frac{3 \, m_Z^4(\varphi,0)}{64 \, \pi^2} \, \left({\rm log} \left[\frac{a_b \, T^2}{\mu^2} \right] - \frac56 \right).
\end{eqnarray}
We choose the renormalization scale $\mu = M_Z$.
The temperature $T_b$ sets the temperature scale at which the phase transition occurs
and dictates the covergence of the high temperature expansion. $T_b$ is a function of
the NP parameters and $m_h$. We find the following for $T_b(m_h)$:
\begin{eqnarray}
\frac{T_b^2(115 \, {\rm GeV})}{(130 \,  {\rm GeV})^2} &\simeq&1 + \frac{(320 \, {\rm GeV})^2}{f^2} \, C_\phi -  \frac{(220 \, {\rm GeV})^2}{f^2} \, \lambda_2, \nn \\
\frac{T_b^2(130 \, {\rm GeV})}{(150 \, {\rm GeV})^2} &\simeq& 1 + \frac{(310 \,{\rm GeV})^2}{f^2} \, C_\phi -  \frac{(190 \, {\rm GeV})^2}{f^2} \, \lambda_2,  \nn \\
\frac{T_b^2(160 \, {\rm GeV})}{(170 \, {\rm GeV})^2} &\simeq& 1 + \frac{(300 \, {\rm GeV})^2}{f^2} \, C_\phi -  \frac{(140 \, {\rm GeV})^2}{f^2} \, \lambda_2, \nn
\end{eqnarray}
where we have rounded to two significant digits and used the zero temperature relationship between $m_h^2$ and 
$\lambda_1$.
We will consider $T_b \sim 100 \,  {\rm GeV}$ in what follows.
Note that  for the region of parameter space where $\lambda_2 \sim \mathcal{O}(1)$ and positive and $ C_{\phi} \sim \mathcal{O}(1)$ and negative the effects of NP can significantly reduce $T_b^2$ and can even in principle 
cause the sign of $T_b^2$ to change. However, as we will show, when this occurs the EWPT is not sufficiently first order for the washout condition to be satisfied. In fact, one can use the above approximate expressions as a quick check of the low energy expression of a PGH model to see if the washout condition is potentially passed.  

\subsubsection{Relation to previous work}\label{past}

Our final potential agrees with some aspects of past studies  
\cite{Zhang:1992fs,Bodeker:2004ws,Delaunay:2007wb,Grojean:2004xa,Noble:2007kk}
although we do find some 
disagreements and our results extend previous investigations in a number of ways. 
The origin of the disagreements and improvements are the following.

We reiterate that our demand
for a reliable perturbative study imposed the power counting $\lambda_1 \lesssim g_2^3$,
thus,  we do not retain the higher order effects of running our parameters. 
We also neglect temperature independent terms in the  
effective potential that are suppressed by $\varphi^2/( 16 \, \pi^2 \,f^2)$. The temperature dependent one loop effects of NP
are retained consistently because this class of terms lead to the critical thermal screening that suppresses the imaginary
part of the effective potential. 
We have introduced a thermal screening due to NP effects that is required when the ring resummation is employed 
with the $\lambda_2$ operator. We note that these
thermal effects are significantly numerically enhanced compared to $\varphi^2/( 16 \, \pi^2 \,f^2)$ effects.

We also reemphasize that we have determined the potential in a canonical low energy effective theory; ie we rescale the Higgs field to
remove the dimension six kinetic terms which introduces the dependence on $C_\phi$ in our effective 
potential. (The dependence on this operator will turn out to be critical when $\lambda_2 \propto \lambda_1$.) Further, we have determined the longitudinal and transverse thermal masses of the gauge boson fields and have used them in our
effective potential.

We have also emphasized that an important feature of the low energy description of PGH models is that the relationship between the Higgs mass and the Higgs self coupling is significantly relaxed in the effective theory, as emphasized in \cite{Grinstein:2007iv}. This effect is essential for perturbative studies of PGBG to be reliable, and is a generic low energy signal of a new strong interaction at a TeV with a PGH. We now turn to determining the condition on the NP parameters and the Higgs mass that allow a first order phase transition to occur while our perturbative study is reliable. 

\section{On the possibility of Pseudo-Goldstone Baryogenesis} \label{phase}

Before turning to the possibility of PGBG, we first review the condition on the parameters in the potential
in the SM for there to be a first order EWPT. This may seem esoteric as if the washout condition is   
passed one knows that the phase transition is first order. However, deriving an analytic constraint on our Wilson coefficients
is useful as it reduces the subsequent region of Wilson coefficients to test for satisfying the washout condition. In addition, our approach in this section 
(and Appendix A) will establish the
region of Wilson coefficients where our perturbative results will be reliable and illustrate how $\lambda_1 <0$ avoids the 
first order phase transition constraint. When this is the case, we derive a further constraint that will ensure our analysis avoids
unreliably concluding PGBG could occur by passing the washout condition when $\lambda_1 < 0$ and the loop expansion is non-perturbative.

\subsection{The first order phase transition condition: SM}
We will emphasize the limitations that the 
non-abelian magnetic mass discussed in  \cite{Espinosa:1992kf,Buchmuller:1993bq} place on the Higgs self coupling in this approach.
This approach determines a limit on the Higgs mass in the SM for a first order EWPT
that reasonably approximates the mass limit determined in lattice investigations. 
Consider the potential of the form of $V_{eff}^{ring}(\varphi,T)$ where one has taken $\lambda_2, C_\phi \rightarrow 0$. We consider the limitations arising from the gauge boson masses. The scalar sector of the theory is known to always 
give a second order phase transition for all values of $\lambda_1$ \cite{Arnold:1992rz,Buchmuller:1993bq}.

One can obtain a necessary condition on the existence of a first order phase transition following \cite{Buchmuller:1993bq}  by
first taking the potential in the simplified form 
\begin{eqnarray}
V(\varphi,T)  &=& \frac{a}{2} \, \left(T^2 - T_b^2 \right) \, \varphi^2 -  \, \sum_i \frac{b_i \, T}{3} \, \left(c_i^2 \, T^2 + \varphi^2 \right)^{3/2}, \nn \\
&+&  \frac{\lambda_1}{8} \, \varphi^4,
\end{eqnarray}
where the constants $a, b_i, c_i$ are all positive. Note that in terms of our perturbative 
couplings $g_{SM}$ we have $b_i \propto g_{SM}^3, a \propto g_{SM}^0,  c_{A} \propto g_{SM}^0$. However, it is important to note that for the longitudinal mass of the $W,Z$ fields the parameters are $c^L_{W}, c^L_{Z} \propto g_{SM}^0$ whereas for the transverse mass $c^T_{W}, c^T_{Z} \propto g_{SM}$ when $\varphi \rightarrow 0$.
Insisting that the temperature is high enough that there is a minimum at the 
origin, one imposes $d^2 \, V(0,T) / d \, \varphi^2 \ge 0$ and one finds that for temperatures
\begin{eqnarray}
T ^2 \ge \tilde{T}_b^2 \equiv \frac{T^2_b}{1 - (\sum_i \, b_i \, c_i)/a},
\end{eqnarray}
there is a minimum at the origin. For there to be a first order phase transition we also require that 
for some $\varphi$ above zero we have another minimum with a potential barrier in between the 
two minima.   For field values just above $\varphi = 0$ and for temperatures above $\tilde{T}_b^2$
the potential increases.  At large $\varphi$ values the potential is dominated by the $\varphi^4$
term and therefore also grows with $\varphi$. For there to be  a second
minimum away from $\varphi=0$ there must be a maximum too. The
condition that there is a  second minimum away from the origin is
weakest when the height of the barrier approaches zero. In this case
the location of the maximum of the barrier moves towards
$\varphi=0$. Hence a necesary condition for the first order phase
transition is that as  $T
\rightarrow \tilde{T}_b$ the derivative of the potential vanishes at
some inifinitesimal value of $\varphi$. Taylor exanding the
derivative, we find the condition
\begin{eqnarray}
0 = \frac{a}{2} \, \left(\frac{\sum_i \, b_i \, c_i \,  T_b^2}{a - \sum_i \, b_i \, c_i}\right) -\frac{\varphi^2}{4} \, \left(\sum_i \, \frac{b_i }{c_i}  -  \lambda_1 \right).
\end{eqnarray}
Recall that $a, b_i, c_i,d_i$ are all positive, $a - \sum_i \, b_i \, c_i >0$ and $T_b^2$ is positive in the SM.
Thus for this equation to have a solution for $\varphi > 0$ one must have 
\begin{eqnarray}
\lambda_1 < \sum_i \, \frac{b_i }{c_i} 
\end{eqnarray}
which is the first order phase transition condition for the SM of \cite{Buchmuller:1993bq}. The condition is dominated by the contribution to the constraint for the transverse $W$ and $Z$ masses which gives for
$\sum_i \,  b_i/c_i$
\begin{eqnarray}\label{NPcond}
\frac{3 \, g_2^2}{2^4 \, \gamma} + \frac{3 \, (g_2 \, \cos[\theta'(T_b)] + g_1 \, \sin[\theta'(T_b)])^4}{2^4 \, \pi \, (3 \, \pi^2 \, g_1^2 \, \sin^2[\theta'(T_b)]/2 + 4 \, g_2^4 \, \gamma \, \cos^2[\theta'(T_b)])^{(1/2)}}. \nn
\end{eqnarray}
Using our approximate results for thermal Weinberg angle, tree level results for $g_1, g_2$ and 
$\gamma= 4.2$ \cite{Heller:1997nqa} we find the phase transition is first order in the SM for a
Higgs mass
\begin{eqnarray}
m_h< 58 \, {\rm GeV}.
\end{eqnarray}
This is consistent with general expectations that the transition is first order in the 
SM if $m_h \lesssim m_W$ and qualitatively agrees with lattice simulations \cite{Kajantie:1996qd,Kajantie:1996mn,Kajantie:1995kf,Csikor:1998eu}. For example \cite{Csikor:1998eu}
finds a first order EWPT for the SM for  $m_H < 72.4 \, \pm \, 1.7 \, {\rm GeV}$.
Thus we consider this condition in the context of new physics to analytically study the relaxation of this 
bound in PGBG.
\subsection{First order phase transition condition: PGBG}
In our PGBG scenario,our effective theory introduces the following changes
in the potential.

\vspace{0.5cm}
\begin{figure*}[tbp]
\centerline{\includegraphics[height = 8.5cm]{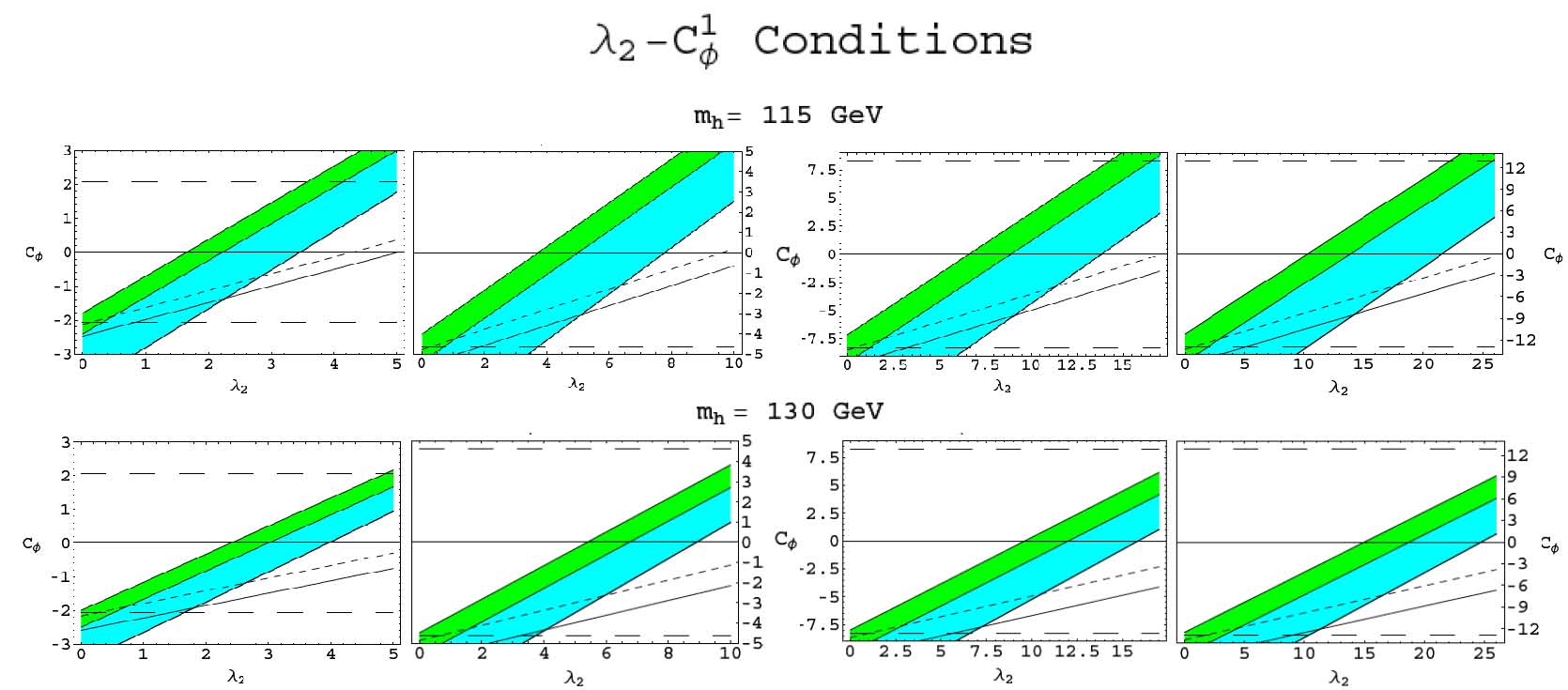}}
\caption{Case $1$ where $\lambda_2$ and $\lambda_1$ are treated as independent. The green ($0 < \overline{\lambda_1}(f,T)  <  5.6 \times 10^{-2}$) and light blue ($-0.2 < \lambda_1  <  0$) regions
satisfy the first order phase transition (and small $\lambda_1$) conditions for Higgs masses of $115 {\rm \, GeV}$ (top) and $130 {\rm \, GeV}$ (bottom).
Also plotted is the condition that $2 v^2 C_\phi/f^2 < 1$ which is the region between the horizontal dashed lines,  Eqn.~(\ref{firstrec}) which is satisfied below the short dashed line and the ascending solid line above which $T_b^2$ is positive. For each Higgs mass we plot the region of allowed Wilson coefficients for a strong decay constant of $f = 500 \, , 750 \, ,1000 \, ,1250 \, {\rm GeV}$ (left to right). The region
that our calculation is self consistent, with a perturbative loop expansion that is under control, and has the signs of $\lambda_1$ and $m^2$ the same as in the SM is the small region in the green band bounded between the ascending solid and short dashed lines. For almost all of the viable parameter space the nature of the EW phase transition is
different than in the SM. For the blue region the potential must be stabilized by the $\lambda_2$ operator.}
\end{figure*}

\vspace{0.5cm}
\begin{figure*}[tbp]
\centerline{\includegraphics[height = 8.5cm]{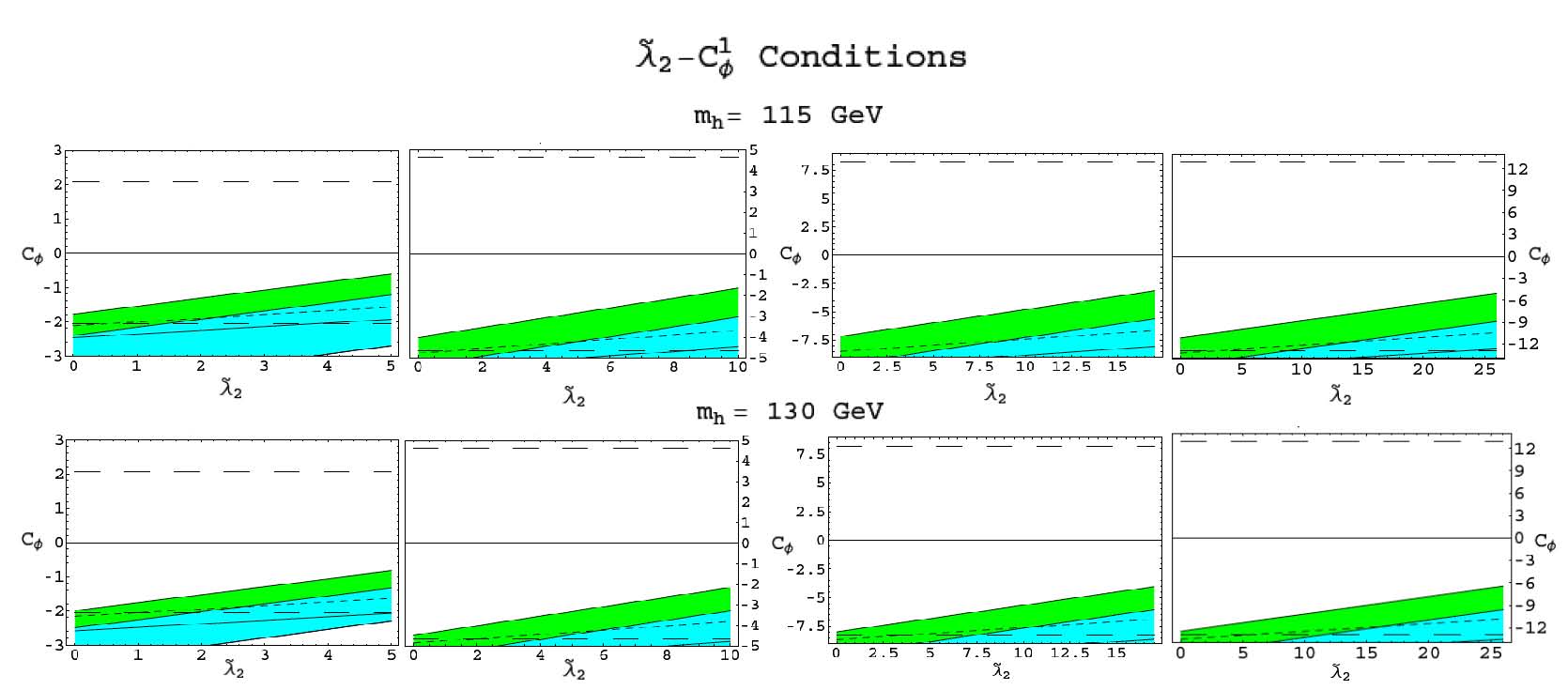}}
\caption{Case $2$ where we plot $\tilde{\lambda_2}=\lambda_2/\lambda_1$. As in Fig. 4,  the green ($0 < \overline{\lambda_1}(f,T)  <  5.6 \times 10^{-2}$) and light blue ($-0.2 < \lambda_1  <  0$) regions
satisfy the first order phase transition (and small $\lambda_1$) conditions for Higgs masses of $115 {\rm \, GeV}$ (top) and $130 {\rm \, GeV}$ (bottom).The lines are the same as in Fig. 4.
For almost all of the parameter space, $\lambda_1$  is positive, however the nature of the EW phase transition is
still quite different than in the SM as we discuss in Section \ref{washout}. }
\end{figure*}

\begin{itemize}
\item [(i)] The constant $a$ is changed by $\lambda_1 \, v^2 \, C_\phi /(12 \, f^2)$. Note that $a$ is still positive
for the range of NP models we will consider as we expanded in  $C_\phi$ which required
$2 \, v^2 \, C_\phi/f^2 < 1$.
\item [(ii)]   The barrier temperature $T_b$ is changed through the change in $a$ and the term
$\lambda_2 \, v^4/(8 \, f^2)$. Demanding the NP effects are such that $m^2$ is still positive gives 
\begin{eqnarray}
\frac{\lambda_1 \, v^2}{2} + \frac{\lambda_2 \, v^4}{8 \, f^2} >0.
\end{eqnarray}
\item [(iii)]  The coefficient of $\varphi^4$ obtains temperature dependence 
and the effecting coupling that will be bounded is shifted by the  
NP Wilson coefficients.
\item [(iv)] The potential now has a $\varphi^6$ term.
\item [(v)] When one relates the Lagrangian density parameters ($m^2,\lambda_1,\lambda_2$) in terms of the physical parameters ($v,m_h^2, \lambda_i^{eff}$)
one must introduce the dependence on $C_{\phi}$ that comes from canonically normalizing the physical Higgs field.
\end{itemize}

The simplified form of the potential is now
\begin{eqnarray}
V(\varphi,T)  &=& \frac{a}{2} \, \left(T^2 - T_b^2 \right) \, \varphi^2 - \sum_i \, \frac{b_i \, T}{3} \, \left(c_i^2 \, T^2 + \varphi^2 \right)^{3/2}, \nn \\
&+&\overline{\lambda_1}(T,f) \, \varphi^4
+ \frac{\lambda_2}{48\, f^2} \, \varphi^6.
\end{eqnarray}

For there to be a second minima for $\varphi > 0$ we now have the condition
\begin{eqnarray}\label{condNP2}
 &\,&\frac{m_h^2}{v^2}  \left(1 + 2 \, C_{\phi} \frac{v^2}{f^2} \right)  -  \lambda_2 \, \frac{v^2}{2 \, f^2} \nn \\   
 &+& \left(\frac{\lambda_2}{3 \, f^2} - \frac{\lambda_1 \, C_\phi}{f^2} \right) \left(\frac{2 \, m_h^2 \, v^2}{m_h^2 + 4 \, B_T \, v^2} \right) < \sum_i \, \frac{b_i }{ c_i}.
\end{eqnarray}
where we use the zero temperature result for $\lambda_1$ and we again neglect the effects of running
this parameter to $\tilde{T}_b$ as it is a higher order effect. 

Let us examine the constraints on the NP Wilson coefficients.
Numerically, the sum is 
\begin{eqnarray}
\sum_i \, \frac{b_i }{c_i}  \simeq 5.6 \times 10^{-2}.
\end{eqnarray}
for $\gamma =4.2$.
When $\lambda_1 < 0$, which can happen in our HEFT, the first order phase transition condition of the SM is evaded.
However, we will still require $|\lambda_1| \lesssim g_2^3$ (which we conservatively take to be $|\lambda_1| \lesssim 0.2$) so that our perturbative investigation has a loop expansion that is under control, see Appendix $\rm A$. This condition and the first order phase transition condition become the important constraint
\begin{eqnarray}
-0.2 \lesssim \frac{ m_h^2}{v^2} \, \left(1 + 2 \, C_{\phi} \frac{v^2}{f^2} \right)  -  \lambda_2 \, \frac{v^2}{2 \, f^2} \lesssim 5.6 \times 10^{-2}.
\end{eqnarray}
Thus we have two inequalities and four unknowns $m_h, C_\phi, \lambda_2, f$.
As shown in Fig. 4 for $m_h \lesssim 160 \, {\rm GeV}$ a first order phase transition can be present if the 
SM is modified with parametrically enhanced dimension six operators. We plot a number of cases where the strong decay scale is in the range dictated by the requirement of enough CP violation for EWB to occur, i.e.  $500 \, {\rm GeV} \lesssim f \lesssim 1000 \,  {\rm GeV}$.  

Recall our cases defined in Section \ref{models}.
In $C1$, the required Wilson coefficient for $\lambda_2$ is $\mathcal{O}(1)$ with $\lambda_2 > 0$ required. One would also expect $\lambda_2 > 0$ for the potential to be stabilized in the presence of these NP terms. For $C_\phi$ the Wilson coefficient can vanish or be $\mathcal{O}(1)$  and 
of either sign. In $C1$, there are large regions of parameter space where the phase transition is first order when $\lambda_1$ is positive or negative as we show in Fig 4.

In $C2$  the important region of constraint for  $\tilde{\lambda}_2$ and  $C_\phi$ is given by
\begin{eqnarray}
-0.2 \lesssim \frac{ m_h^2}{v^2} \, \left(1 + 2 \, C_{\phi} \frac{v^2}{f^2}  -  \tilde{\lambda}_2 \, \frac{v^2}{2 \, f^2} \right) \lesssim 5.6 \times 10^{-2}.
\end{eqnarray}
In $C2$, for almost all of the parameter space where the phase transition is first order, $\lambda_1$ is positive in our HEFT as we show in Fig 5.
When $\lambda_1>0$ in our HEFT,  we find that so long as $\tilde{\lambda}_2$ is positive and $C_\phi$ is  $\mathcal{O}(1)$ and negative
the phase transition can be first order. We do note however, that the models discussed in Section \ref{models}
all have $C_\phi$ Wilson coefficients that are $\mathcal{O}(1)$ and {\it positive}.
We now turn to determining the critical washout condition in our PGBG scenario. 

\section{Washout Condition} \label{washout}
When considering the washout condition it is best to have a picture of the phase transition in mind. We
plot $V_{eff}^{ring}$ when the parameters in the Lagrangian density are such that a 
first order phase transition is possible for $m_h = 120 \, {\rm GeV}$ in the cases where $\overline{\lambda_1}(T,f) < 0$
and $\overline{\lambda_1}(T,f) > 0$ and for the SM in Fig. 6.

\vspace{0.5cm}
\begin{figure*}[tbp]
\centerline{\includegraphics[height = 13cm]{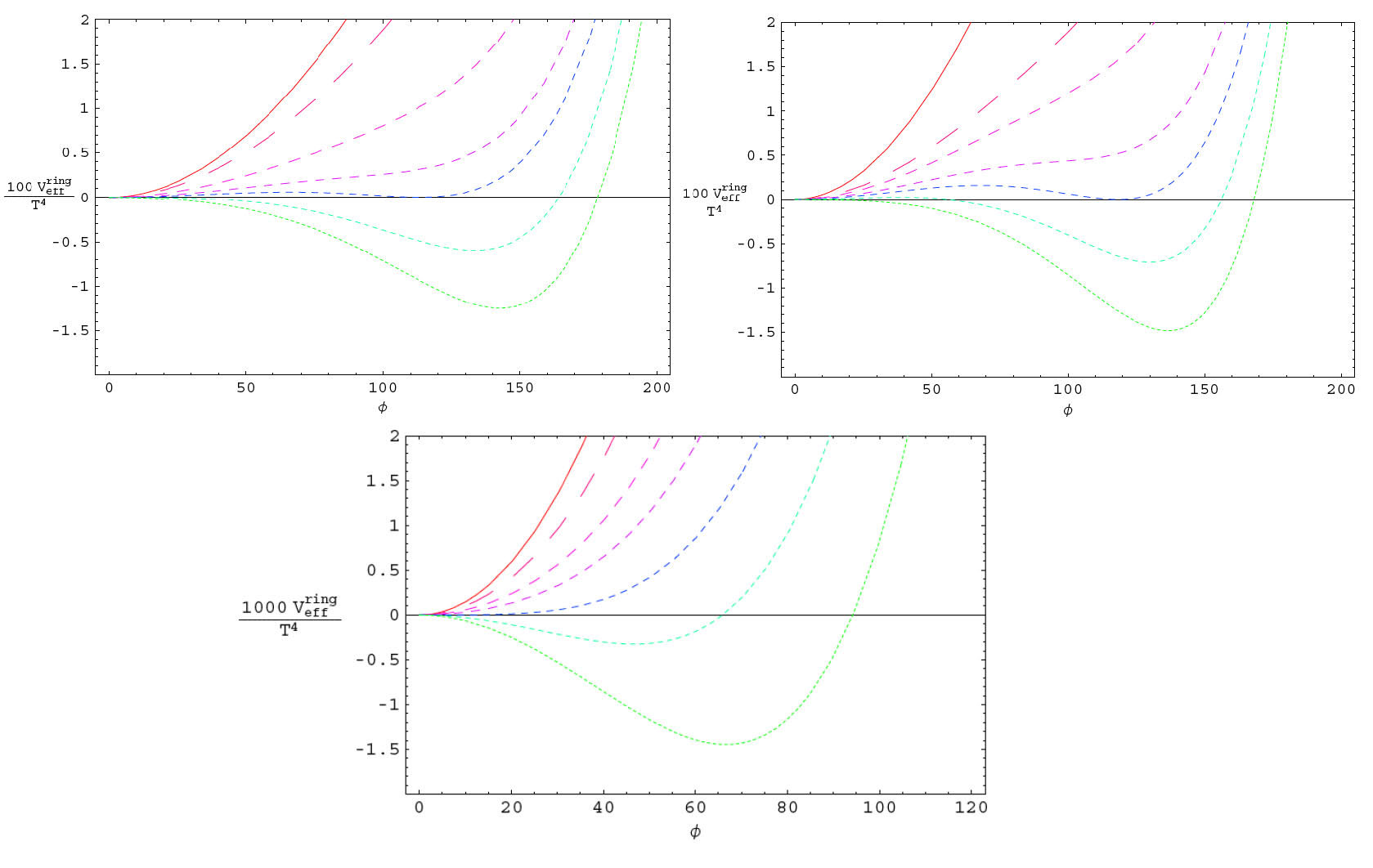}}
\caption{ The temperature dependence of the EW phase transition in a number of cases when $m_h = 120 \, {\rm GeV}$.
As the dashes get shorter and as the colour decends in hue the universe is cooling down. Comparison of the three graphs clearly illustrates the sensitivity of the EW phase transition to the low energy expression of a new strong interaction with a TeV mass scale and the influence of the $C_{\phi}$ operator. The potentials are normalized to zero at the origin.
Top Left:The decay constant is $f = 700  \, {\rm GeV}$ and $\lambda_2 = 4$, $C_{\phi} = 0$. In this case $\lambda_1<0$. Temperatures plotted = ($115, \, 110, \, 105, \, 103, \, T_c = 102.2, \,  101, \, 100$) {\rm GeV} and $\varphi_c = 113.4 \, {\rm GeV}$ so that $\varphi_c/T_c = 1.11$. 
Top Right: The decay constant is $f = 700  \, {\rm GeV}$ and $\lambda_2 =2$, $C_{\phi} = -2$. In this case $\lambda_1>0$. 
Temperatures plotted = ($80, \, 75,  \, 74, \, 73, \, T_c = 72.5, \,  72, \, 71.5$) {\rm GeV} and $\varphi_c = 118.1 \, {\rm GeV}$ so that $\varphi_c/T_c = 1.63$.
Bottom: The SM for comparison. In this case $\lambda_1>0$. Temperatures plotted = ($150, \, 145, \, 141, \,139, \,  Tc = 136.9, \,  135, \, 133$) {\rm GeV} and $\varphi_c = 0.28 \, {\rm GeV}$ so that $\varphi_c/T_c = 2.1 \times 10^{-3}$.
Note that as $\varphi \rightarrow 0$ formally the loop expansion breaks down and thus the behavior of the graphs as 
$\varphi \rightarrow 0$ is not reliably determined in perturbation theory but formally the normalized potential must vanish.}
\end{figure*}

A first order phase transition proceeds through the nucleation of bubbles where $ \varphi  > 0$ inside the bubble, $\varphi  = 0$ outside the bubble and the expectation value of the Higgs changes rapidly as one goes through the bubble wall.
As the universe cools down, when the phase transition occurs eventually a critical temperature $T_c$ is reached where 
the high temperature minima at the origin and the minima at $\varphi_c$ are degenerate.
The conditions defining $\varphi_c,T_c$ are
\begin{eqnarray}
V_{eff}(\varphi_c, T_c)  &=& V_{eff}(0, T_c), \nn \\
\frac{\partial \, V_{eff}( \varphi_c,T_c)}{\partial \, \varphi} &=& 0,
\end{eqnarray} 
and correspond to the blue line in Fig 6.

\begin{figure*}[tbp]
\centerline{\includegraphics[height = 14cm]{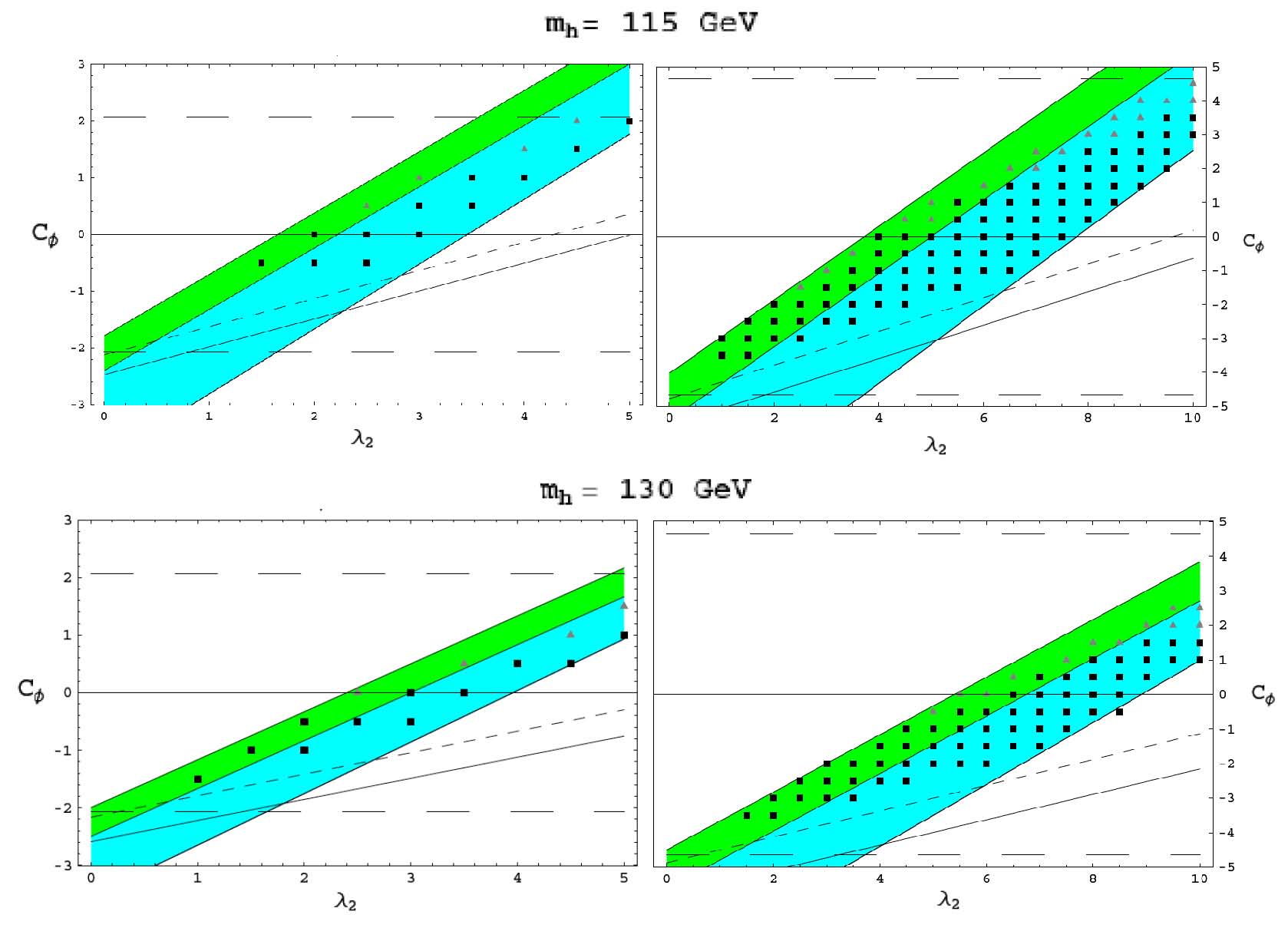}}
\caption{The overlay of the washout condition and our phase transition condition when $\lambda_2$and $\lambda_1$ are independent. The lines are defined in Fig.4 and the decay constant of 
scale is  $f = 500 \, {\rm GeV}$ (left) and  $750 \, {\rm GeV}$ (right). The black square indicates that the stronger washout condition $\varphi_c/T_c \geq 1.3$ is passed, the grey triangle indicates that only the weaker washout condition $\varphi_c/T_c \geq 1.0$ is passed.}
\end{figure*}

\begin{figure*}[tbp]
\centerline{\includegraphics[height = 14cm]{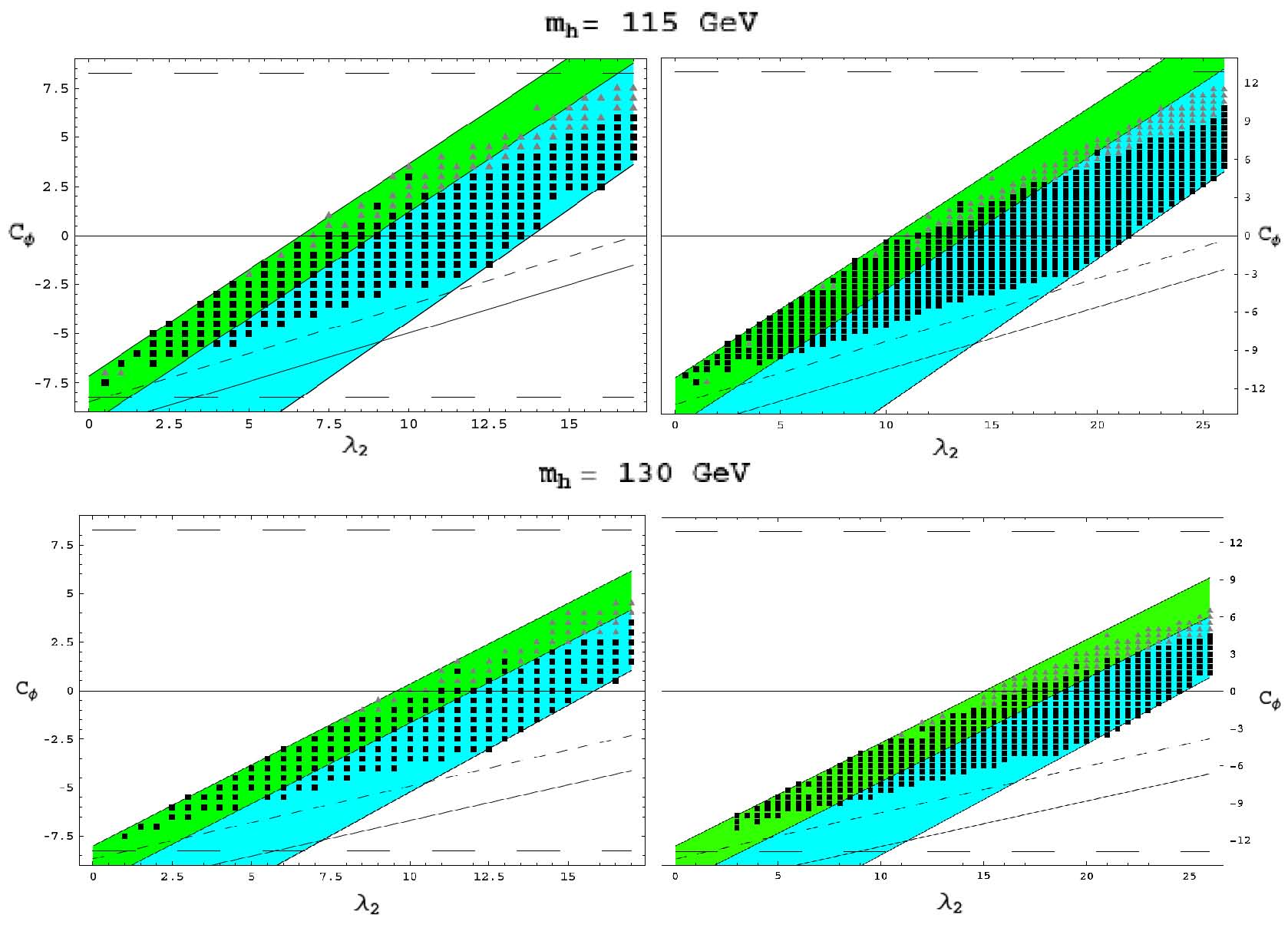}}
\caption{The overlay of the washout condition and our phase transition condition when $\lambda_2$and $\lambda_1$ are independent. The lines are defined in Fig.4 and the decay constant of 
scale is  $f = 1000 \, {\rm GeV}$ (left) and  $1250 \, {\rm GeV}$ (right). The black square indicates that the stronger washout condition $\varphi_c/T_c \geq 1.3$ is passed, the grey triangle indicates that only the weaker washout condition $\varphi_c/T_c \geq 1.0$ is passed. As the scale $f$ grows, the size of the required Wilson coefficient for $\lambda_2$ grows rapidly.}
\end{figure*}

\begin{figure*}[tbp]
\centerline{\includegraphics[height = 14cm]{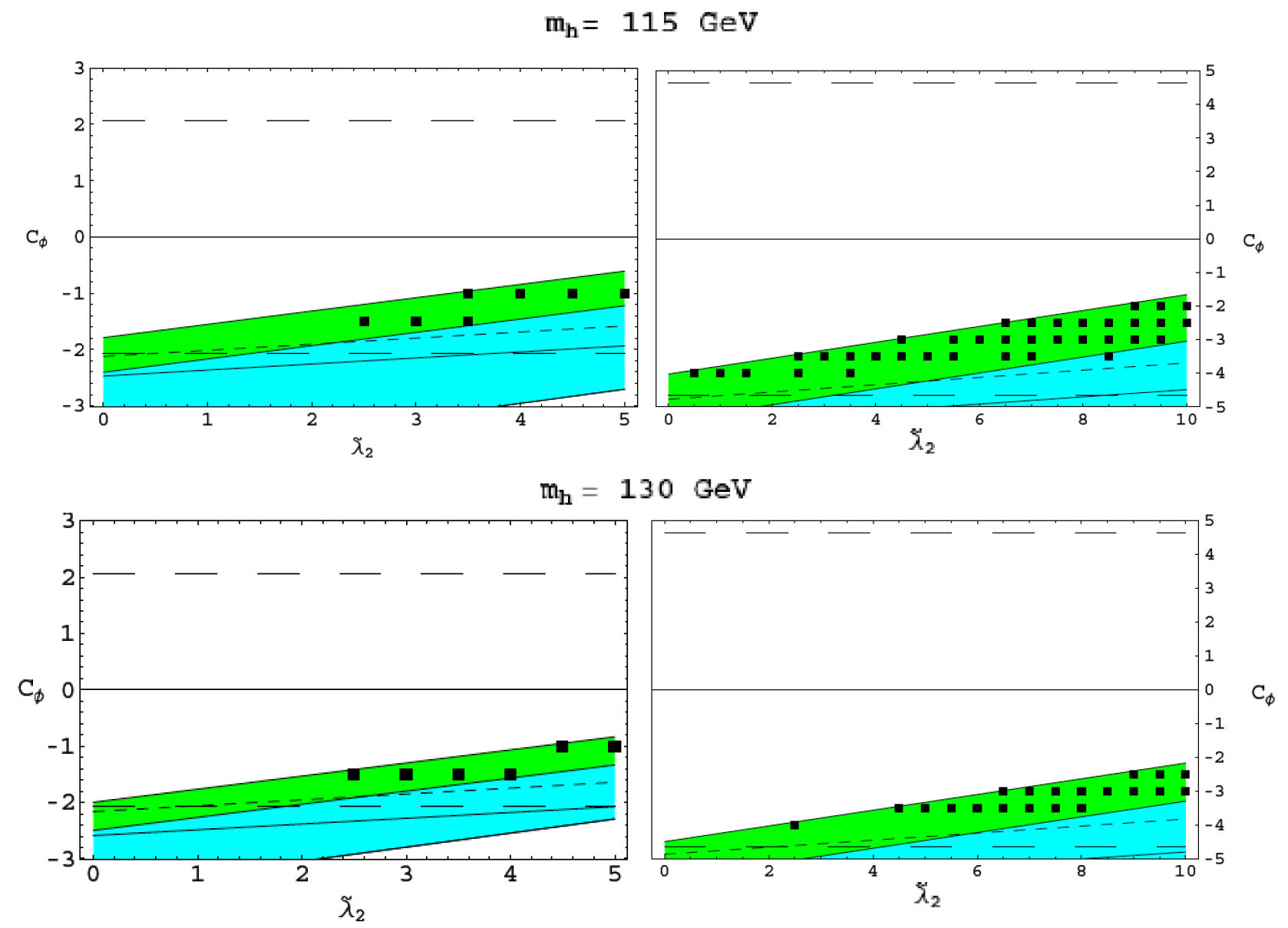}}
\caption{The overlay of the washout condition and our phase transition condition when $\lambda_2$ is proportional to $\lambda_1$. The lines are defined in Fig.4 and the symbols in Fig. 8. The decay constant 
scale is  $f = 500 \, {\rm GeV}$ (left) and  $750 \, {\rm GeV}$ (right).}
\end{figure*}

\begin{figure*}[tbp]
\centerline{\includegraphics[height = 14cm]{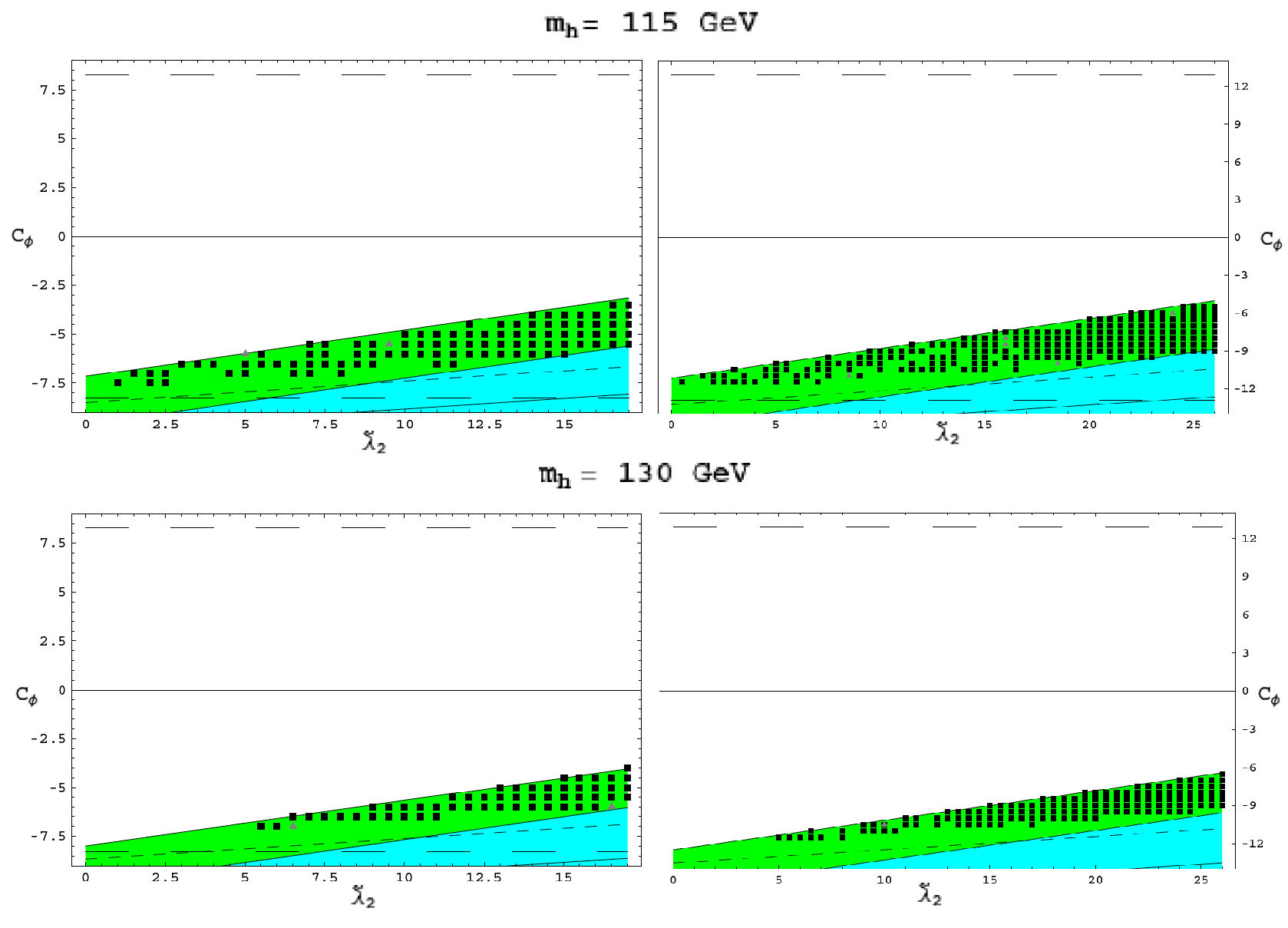}}
\caption{The overlay of the washout condition and our phase transition condition when $\lambda_2$ is proportional to $\lambda_1$. The lines are defined in Fig.4 and the symbols in Fig. 8. The decay constant 
scale is  $f = 1000 \, {\rm GeV}$ (left) and  $1250 \, {\rm GeV}$ (right).}
\end{figure*}

We wish to solve for $\varphi_c$ and $T_c$ as the washout condition 
must be satisfied for the phase transition to be sufficiently first order. Sufficiently first order is defined as the condition
discussed in  Sec. \ref{sectwo}
\begin{eqnarray}\label{wout}
\varphi_c/T_c \geq b, 
\end{eqnarray} 
with $1 \lesssim b \lesssim 1.3$. Satisfying the washout condition \cite{Dine:1991ck}
guarantees that once Baryogenesis has taken place outside the bubble wall, as the bubble expands and 
envelops the produced Baryon number, the remaining sphaleron induced $\rm B+L$ violating Boltzman fluctuations inside the bubble do not erase the produced Baryon number.  The sensitivity of the right hand side of Eqn.~(\ref{wout}) to the $\varphi^6$ term was examined in \cite{Grojean:2004xa} and found to be a percent level effect that we neglect.

As our effective potential is quite complicated, we solve for $\varphi_c,T_c$ numerically. The procedure we use is to first translate $V_{eff}^{ring}[v,\lambda_1,\lambda_2,C_{\phi},f;T,\varphi] $ to
\begin{eqnarray}
V_{eff}^{ring}[v,m_h^2,\lambda_2,C_{\phi},f;T,\varphi] 
\end{eqnarray} 
using our zero temperature definition of $\lambda_1$ while neglecting the effects of running. We then 
choose a $m_h^2,f$ and numerically solve for $\varphi_c$ and $T_c$ by scanning the allowed region of $\lambda_2, C_{\phi}$ parameter space determined in Fig. 4 and 5. Our results are reported in 
Fig. 7-10.

We find that $T_b^2 >0$ when the washout condition is passed. We note that there is a region where 
$T_b^2 >0$ roughly parallel to the  $T_b^2 = 0$ line where the washout condition is not passed. We find empirically that the following constraint equation determines this region where the washout condition is not passed and $T_b^2 >0$,
\begin{eqnarray}\label{firstrec}
{\large{|}} a \, T_b^2 &-& \sum_i \, b_i \, T_b \, \sqrt{c_i^2 \, T_b^2 + v^2} \\
&+& \frac{T_b^2 \, v^2}{f^2} \, \left(\frac{\lambda_2}{6} - \frac{4 \, \lambda_1 \, C_\phi}{9}\right)
+ \frac{\lambda_2 \, v^4}{8 \, f^2} {\large{|}} <  {\large{|}}\frac{\lambda_1 \, v^2}{2}{\large{|}}\nn
\end{eqnarray}

This equation is inspired by the fact that  it is known that the SM with a Higgs mass in the region we consider $114.4 \, {\rm GeV} \lesssim m_h \lesssim 160 \, {\rm GeV}$ does not pass the washout condition, thus the relationship between the $m^2$ and $\lambda_1$ parameters need to be significantly effected in order to satisfy our washout condition. The equation characterizes the relationship between $\lambda_1$ and $m^2$ as $T \rightarrow T_b$.  When the washout condition is not passed and this equation is satisfied, the critical signs of $m^2$ and $\lambda_1$ are the same (although both are negative in C$1$ unlike in the SM). 
In the region of parameter space dictated by this equation and the $T_b^2 >0$ condition, the washout condition is not passed.

This condition can be understood by the following approximation in case 1. Recall the zero temperature 
minimization condition, Eqn.~(\ref{min}) when $\varphi = v$. When $m^2 = 0$ this equation dictates
\begin{eqnarray}
\lambda_1 = - \frac{v^2}{4 \, f^2} \, \lambda_2.
\end{eqnarray}
Substituting this result in Eqn.~(\ref{higgsmass}) we obtain the constraint
\begin{eqnarray}
C_{\phi} = \frac{\lambda_2}{8} \, \frac{v^2}{m_h^2} - \frac{f^2}{2 \, v^2},
\end{eqnarray}
that reasonably approximates the plotted Eqn.~(\ref{firstrec}).
When the 
washout condition is satisfied in PGBG, $T_b^2>0$ and the relationship between $m^2$ and $\lambda_1$ 
must be significantly effected in the sense that Eqn.~(\ref{firstrec}) is not satisfied.  There is a further constraint on PGBG due to the effect that an expanding universe has on the possibility of the bubble formation. The results of \cite{Delaunay:2007wb} indicate that for the case  $C_{\phi} =0$ the supercooling effect due to the expansion of the universe is a small shift in $T_c$ for most of the relevant parameter space. The temperatures for most of the parameter space above are $\sim 100 \, {\rm GeV}$. 

\section{Conclusions}\label{conc}
We have shown how an effective theory of the SM Higgs that would be the low energy description of a PGH can address all of the problems of EW scale SM baryogenesis. Our results indicate that PGH models with Wilson coefficients $\lambda_2$ and $C_{\phi}$ that are $\mathcal{O}(1)$
and a strong decay scale $f$ in the range ($500 \, {\rm Gev}, 1 \, {\rm TeV}$) may 
successfully account for the origin of the baryon-antibaryon asymmetry of the universe.

The PGBG scenario is falsifiable and should be ruled in or out as the possible origin of the Baryon asymmetry of the universe
in the next few years of experiments.
Let us consider the experimental path that could find evidence for PGBG being the origin of the
baryon-antibaryon asymetry of the universe. 

If the Higgs self coupling can be determined through the process $gg \rightarrow h \, h$ \cite{Dawson:1998py,Grinstein:2007iv} at LHC and it deviates from the SM value dictated by the determined Higgs mass, our results indicate that one should start to seriously consider 
PGBG. A large effect on the relationship between $\lambda_1$ and $m_h$ in the effective theory is absolutely required. If this is established and ideally new resonances of a new strong interaction  were discovered then 
the possibility is seriously raised that PGBG may be the origin of a significant amount of baryon-antibaryon asymmetry in the universe. Unfortunately the limited kinematic reach of LHC means that 
new strong interaction states could easily be elusive at LHC. Indirect signals of a new strong interaction such as a growth in the longitudinal gauge boson scattering amplitudes despite the presence of a light Higgs \cite{Giudice:2007fh} are possibly the best
that can be achieved experimentally. If strong interaction states avoid detection due to LHC's limited reach,
the large effects of NP in the Higgs sector 
in this scenario allows one to have some reasonable hope of interesting signals of NP in the properties of the Higgs.

In conjuction to these LHC results, PGBG also requires that electric dipole moment (EDM) experiments also find evidence for non-SM CP violation. If such a set of discoveries are made, one will actually be
able to conclude that that PGBG is the likely source of the baryon-antibaryon asymmetry in the 
universe.\footnote{Leptogenesis with new sources of CP violation in the lepton sector would not 
induce such large effects on EDMs. EDMs do not violate lepton number and $\delta \, d_e \sim G_f^2 \, m_e \, m\nu^2$ \cite{Huber:2006ri,Pospelov:2005pr}.} 

This scenario has a number of interesting features that increase its viability. 
The $SU_C(2)$ operators that are parametrically enhanced in the Higgs sector (and only exactly these operators) are exactly 
the operators that need to be sizable in our HEFT for PGBG to occur. These operators are not constrained by EWPD to be small. 
An interesting feature of PGBG is the coincidence in the required strong decay scale $f$.
The same range of scales is required for   
the SM to be supplemented with enough CP violating effects and the EW phase transition to be sufficiently first order. 

If the Higgs is found at LHC and if it is a pseudo-goldstone Higgs, experiment could soon inform us if 
EW Pseudo Goldstone Baryogenesis is the origin of the baryon-antibaryon asymmetry of the universe.

\begin{acknowledgments}
Work supported in part by the US Department of Energy under contract DE-FG03-97ER40546.
\end{acknowledgments}

\appendix

\section{Constraints for Reliable Perturbative Studies of NP and the EW phase transition}\label{reliable}

We have emphasized the need to have a loop expansion under control in thermal field theory calculations of the 
EWPT in our HEFT. We digress for a moment to give some more detail on  why this consideration is so important. The concern about the convergence of perturbation theory is more urgent in investigations of the
EWPT. As discussed in Section \ref{finiteT}, finite temperature effects are known to cause the loop expansion to break down for sufficiently high temperatures leading to high temperature symmetry restoration.
In the SM, even with ring improvement, the loop expansion is still a poor expansion if the scalar doublet quartic self coupling
$\lambda_1$ is large \cite{Arnold:1992rz,Arnold:1994bp,Buchmuller:1993bq}. 
Once we employ ring resummation to absorb the thermal mass terms that scale as $\mathcal{O}(\lambda \, T^2)$ and $\mathcal{O}(g^2 \, T^2)$, the remaining loops are dominated by momenta of the order of their mass scale and the loop expansion parameters are dictated by $\lambda_1 \, T/m_{eff}$ and $g_{SM}^2 \, T/m_{eff}$ \cite{Weinberg:1974hy,Arnold:1992rz}. These loop expansion parameters place a constraint on $\lambda_1$ for perturbative studies to be reliable. 

As an example to clarify the issue, consider the ring improved potential of a pure scalar theory. This potential appears to give a first order phase transition
at leading order in the ring improved loop expansion. However this conclusion is incorrect. A pure scalar theory
is well known to undergo only a second order phase transition. This incorrect conclusion is reached as the loop expansion parameter is order one near the phase transition \cite{Arnold:1992rz}. For the pure scalar theory the loop expansion parameter is  $\lambda \, T / m_{eff}$ and $m_{eff} \sim \lambda \, T$. This clearly illustrates the need to 
insist that perturbative studies take note of the nature of the expansion parameter and ensure that it is less than one.

Now consider the (lower order) simplified classical potential inspired by our effective potential of the form
\cite{Arnold:1992rz}
\begin{eqnarray}
V(\varphi,T) &=& \frac{1}{2} \, \left(a \, g^2 T^2 - m^2 \right) \, \varphi^2+ \frac{\lambda_1}{8} \, \varphi^4, \nn \\
&\,&  -  \, \sum_i \frac{b_i \, T}{3} \, \left(c_i^2 \, T^2 + \varphi^2 \right)^{3/2}. 
\end{eqnarray}
The phase transition occurs when all terms in the potential are approximately the same size. When this occurs one finds
\begin{eqnarray}
\varphi_c \sim \frac{g^3}{\lambda_1} \, T_c, \quad  \left(a \, g^2 \, T_c^2 - m^2 \right) \sim \frac{g^6}{\lambda_1} \, T_c^2.
\end{eqnarray}
For this potential, the transverse vector loop expansion parameter (subloops with $M_W^T(\varphi,T)$ vectors running in them) is given by
\begin{eqnarray}
\frac{g^2 \, T_c}{M_W^T(\varphi,T)} \sim \frac{\lambda_1}{g_2^2}. 
\end{eqnarray}
This is why we insist that for the loop expansion to be under control one must have 
$\lambda_1 \sim g_2^3$ \cite{Arnold:1992rz,Arnold:1994bp,Buchmuller:1993bq}.Perturbative studies that do not 
take this constraint into account run the risk of obtaining unreliable conclusions. For Higgs masses above $m_h > 115 \, {\rm GeV}$, perturbative studies of the EWPT of the SM are unreliable for this reason.

In our perturbative investigation of our PGBG scenario we must insist that  {\it{the values of  $\hat{m}_h, f$ and the Wilson coefficients $\lambda_2, C_{\phi}$ dictate that the loop expansion is under control.}}  This is not a significant fine tuning for $m_h \lesssim 160 \, {\rm GeV}$. Due to the parametrically enhanced NP effects on the relationship between the Higgs mass and the Lagrangian parameter $\lambda_1$ in our effective theory. For PGH models in case $C1$ this suppression of the Higgs self coupling naturally occurs when  $\lambda_2 \sim \mathcal{O}(1)$ and positive and $ C_{\phi} \sim \mathcal{O}(1)$.  This suppression of the Higgs self coupling in our HEFT also tends to make the phase transition first order while improving the justification of perturbative studies. $\lambda_2>0$ is also desired so that NP stabilizes the Higgs potential when $\lambda_1 < 0$.\footnote{Lattice simulations could relax this constraint on $\lambda_1$ while investigating the nature of the electroweak phase transition in PGH scenarios.}  

Some past studies have allowed the Higgs mass to be $m_h \gg 160 \, {\rm GeV}$
when considering the effect of NP \cite{Bodeker:2004ws,Delaunay:2007wb,Grojean:2004xa} and have not taken this constraint on $\lambda_1$ into account.  As the Higgs mass increases the $\lambda_2$ Wilson coefficient must become 
rather large for $\lambda_1$ to remain small.  One should also note that Higgs masses above 
$160 \, {\rm GeV}$ one has a poorly behaved high temperature expansion. For these reasons we restrict our investigation to $m_h \lesssim 160 \, {\rm GeV}$.

This reasoning also gives a constraint on $\lambda_2$ for a reliable perturbative investigation.
Consider the non renormalizable potential of the form 
\begin{eqnarray}
V_{eff}^{ring}(\varphi, T) &=& \frac{a}{2} \left(T^2 - T_b^2 \right) \, \varphi^2 + \frac{\lambda_2}{48 \, f^2} \, \varphi^6
+ \overline{\lambda_1}(T,f) \, \varphi^4, \nn \\
&\,&  -  \, \sum_i \frac{b_i \, T}{3} \, \left(c_i^2 \, T^2 + \varphi^2 \right)^{3/2}. 
\end{eqnarray}
Again, the phase transition occurs when all terms in the potential are approximately the same size. When this occurs one again finds the constraint $\lambda_1 \sim g_2^3$ is appropriate and 
we have the additional condition 
\begin{eqnarray}
\lambda_2 \sim \lambda_1 \, \frac{f^2}{T_c^2}.
\end{eqnarray}
For the decay constant scale $f$ and critical temperatures $T_c$ of interest one finds
that 
\begin{eqnarray}
\lambda_2 \sim g_2^3 \, \frac{f^2}{T_c^2} \sim \mathcal{O}(1),
\end{eqnarray}
which is consistent with the values of $\lambda_2$ we find are required for a first order phase transition in
the context of NP.

When $\lambda_1 < 0$ in our low energy PGH Lagrangian the first order phase transition condition of the SM is evaded.
However, we will still require $|\lambda_1| \lesssim g_2^3$ (which we conservatively take to be $|\lambda_1| \lesssim 0.2$) so that our perturbative investigation has a loop expansion that is under control. This condition and the first order phase transition condition become the important constraint
\begin{eqnarray}
-0.2 \lesssim \frac{ m_h^2}{v^2} \, \left(1 + 2 \, C_{\phi} \frac{v^2}{f^2} \right)  -  \lambda_2 \, \frac{v^2}{2 \, f^2} \lesssim 5.6 \times 10^{-2}.
\end{eqnarray}
This equation can be satisfied for large regions of $C_{\phi}, \lambda_2$ parameter space when $m_h \ge 115 \, {\rm GeV}$. The upper bound on this constraint equation has some finite temperature effects that we discuss in Section \ref{phase}. The requirement of $|\lambda_1| \sim g_2^3$ that is appropriate for $\lambda_1 <0$ is purely a requirement for a loop expansion that is under control and independent of temperature. 

\section{High Temperature expansions}

In calculating the  $W, \, Z, \, A$
 thermal mass terms that influence the nature of the EWPT
we employ high temperature expansions.
All temperature dependent loop integrals can be decomposed in terms of a basic integral

\begin{eqnarray}
J_{\pm}(y^2) \equiv \int_0^\infty d x \, x^2 \, \log \left[{1 \mp \, \exp{\left(- \sqrt{x^2 + y^2} \right)}} \right],
\end{eqnarray}

\noindent{where $y^2_i = m^2_i/T^2$. The derivatives of this integral we denote}
\begin{eqnarray}
I_{\pm}(y^2) &=&  2 \, \left[ \frac{d \, J_{\pm}(y^2)}{d \, y^2}\right], \\
K_{\pm} (y^2) &=& \frac{d \, I_{\pm}(y^2)}{d \, y^2}.
\end{eqnarray}

Expressing our results in terms of these integrals allows one to easily improve the propagators self consistently
with the determined thermal masses.
We employ high temperature expansions for these integrals obtained by
taking derivatives of the following expansions 

\begin{eqnarray}
J_{+}(y^2) &=& \frac{\pi^2 \, y^2}{12} - \frac{\pi \, (y^2)^{3/2}}{6} - \frac{y^4}{32} \, \log \left[\frac{y^2}{a_b} \right] + \mathcal{O} \left(\frac{y^{3+ n}}{2^n \, \pi^n}\right),  \nn \\
J_{-}(y^2) &=&- \frac{\pi^2 \, y^2}{24}  - \frac{y^4}{32} \, \log \left[\frac{y^2}{a_f} \right] + \mathcal{O} \left(\frac{y^{3+ n}}{2^n \, \pi^n}\right),
\end{eqnarray}

\noindent{where $ n = 2,3 \ldots$ and $\log a_b = 5.408$, \, \, $\log a_f = 2.635$. We also find it convenient to define the following functions which are the results of common loop integrals 
\begin{eqnarray}
F_1(y_a^2,y_b^2) &=& \frac{1}{2} \frac{\left(y_a^2 I_{+}(y_a^2) -  y_b^2 I_{+}(y_b^2) \right)}{y_a^2 - y_b^2} + F_6(y_a^2,y_b^2) \nn \\
F_2(y_a^2) &=& \frac{1}{2} \left[y_a^2 \, K_{+} (y_a^2)  - \frac{1}{2} \, I_{+}(y_a^2) \right] \nn \\
F_3(y_a^2,y_b^2)  &=&  F_1(y_a^2,y_b^2) + F_2(y_a^2), \nn \\
F_4(y_a^2,y_b^2) &=& I_{+}(y_a^2)+ 2 \, I_{+}(y_b^2)  + \frac{3}{y_a^2} \left(J_{+}(0)- J_{+}(y_a^2) \right), \nn \\
F_5(y_a^2,y_b^2) &=& \frac{3 \, \varphi^2}{2 \, \pi^2 \left(y_a^2 - y_b^2\right)} \left[\frac{J_{+}(y_a^2)}{y_a^2} - \frac{J_{+}(y_b^2)}{y_b^2} \right] \nn \\
F_6(y_a^2,y_b^2) &=& -  \frac{3}{2} \left( \frac{\left(J_{+}(y_a^2)- J_{+}(y_b^2) \right)}{y_a^2 - y_b^2} \right) \nn \\
F_7(y_a^2,y_b^2) &=&  I_{+}(y_a^2)+ 2 \, I_{+}(y_b^2)  + \frac{3}{2 \, y_a^2} \left(J_{+}(y_a^2)- J_{+}(0) \right), \nn \\
F_8(y_a^2,y_b^2) &=&  \frac{ \varphi^2}{3 \, \pi^2 \left(y_a^2 - y_b^2\right)} \, \left(I_{+}(y_a^2) -  I_{+}(y_b^2) \right),
\end{eqnarray}
We express our results in terms of these functions to allow our results to be used, if desired, when the high temperature expansion
is not employed.

\subsection{Thermal masses: Scalars }
 
As a simple example of the techniques employed to determine the thermal mass basis for the gauge bosons 
we now obtain the thermal masses of the Higgs and goldstone boson fields in the high temperature limit.  
We also improve on past results by using the reasoning behind the one loop ring resummation to consistently employ ring resummation when
a  $\lambda_2$ operator is present. This will reduce the imaginary part of the effective potential due to the scalar masses.
To obtain thermal mass contributions appropriate to shift the mass in ring resummation, one sets the external momenta $(k_0, \bf{k})$ to zero by setting $k_0 = 0$ and taking the limit ${\bf k^2} \rightarrow 0$. 
The diagrams to determine are thermal loops given by Fig. 11. We calculate in the $W^I, B$ basis for the gauge
bosons as we are interested in leading order $T^2$ effects for the scalars to illustrate the modification of the results due to the presence of NP.\footnote{Note that we use the results of the one loop gap calculations for the scalars (neglecting NP) in \cite{Buchmuller:1993bq} when we consider the one loop gap equation results of the vector bosons to be consistent in the vector boson section.} . 
\begin{figure}[bh] \label{scalartad}
\centerline{\scalebox{0.7}{\includegraphics{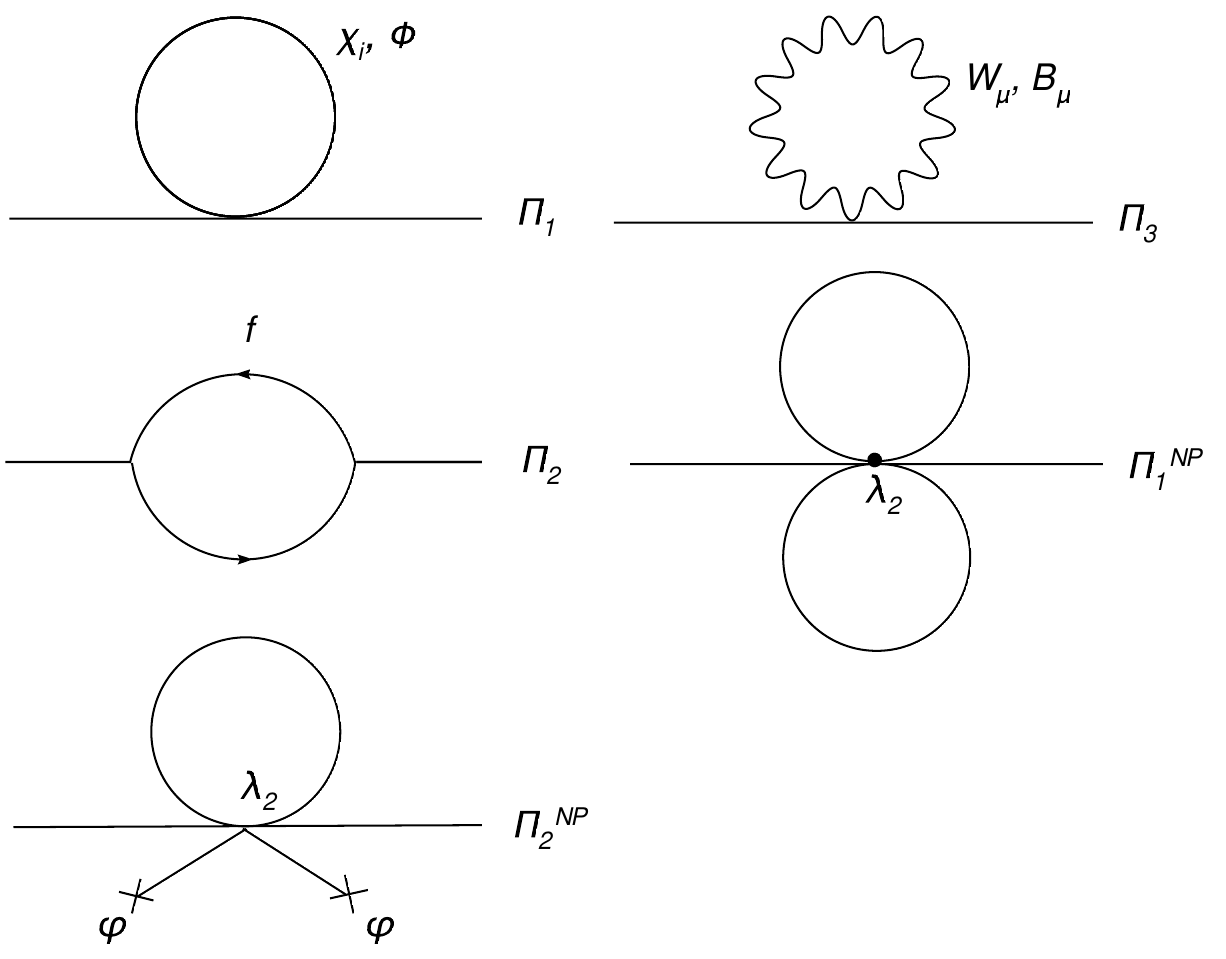}}}
\caption{One loop diagrams that contribute to the scalar thermal masses.}
\end{figure}

The results for the finite temperature contributions to the scalar self energies for the Higgs are
\begin{eqnarray}
\Pi_1 &=&  \frac{3 \, \lambda_1 \, T^2}{4 \, \pi^2} \left( I_{+} (y^2_h)(1 - 4 \,  C_{\phi} ^1\, \frac{\varphi^2}{f^2}) \right), \\
&\,&  +  \frac{3 \, \lambda_1 \, T^2}{4 \, \pi^2} \left( I_{+} (y^2_{\chi}) (1 - 2 \,  C_{\phi} ^1\, \frac{\varphi^2}{f^2})   \right),  \nn\\
\Pi_2 &=& - \frac{6\, m^2_t \, T^2}{\varphi^2 \, \pi^2} \left( y_t^2 \, K_{-} (y^2_t)  + I_{-} (y^2_{t}) \right) (1 - 2 \,  C_{\phi} ^1\, \frac{\varphi^2}{f^2}) , \nn \\
\Pi_3 &=&  \frac{3 \, T^2}{8 \, \pi^2} \left( g_1^2 \, I_{+} (y^2_B) + 3 \, g_2^2 \,  I_{+} (y^2_{W})  \right) (1 - 2 \,  C_{\phi} ^1\, \frac{\varphi^2}{f^2}).  \nn \\
\Pi_1^{NP} &=& \frac{ \lambda_2}{f^2} \left[ \frac{3 \, \lambda_1 \, T^2}{2 \, \pi^2} \, \left( I_{+} (y^2_h) + I_{+} (y^2_\chi) \right) \right]^2, \nn \\
\Pi_2^{NP} &=& \frac{9 \, \lambda_1 \, T^2}{2 \, \pi^2} \left( I_{+} (y^2_h) + I_{+} (y^2_\chi) \right) \, \left( \frac{\varphi^2 \, \lambda_2}{f^2} \right), \nn
\end{eqnarray}
where we have included the necessary rescalings of the kinetic sector to have a canonical low energy theory.
Utilizing the high temperature expansion to expand in each case to leading order we find 
\begin{eqnarray}
\Pi_{h}(\varphi,T) &=& \frac{T^2 \, \lambda_1}{4} \left(1 - 3 \, C_{\phi} ^1 \frac{\varphi^2}{f^2} \right)
+ \frac{\varphi^2 \, T^2}{2 \, f^2} \, \lambda_2, \nn \\
&+& T^2 \, B_T \,   \left(1 - 4 \, C_{\phi} ^1 \frac{\varphi^2}{f^2} \right) + \frac{3 \, T^4 \, \lambda_2}{4 \, f^2}, \nn \\
\Pi_{\chi}(\varphi,T) &=&  \frac{T^2 \, \lambda_1}{4} \left(1 -  C_{\phi} ^1 \frac{\varphi^2}{f^2} \right)
+ \frac{\varphi^2 \, T^2}{2 \, f^2} \, \lambda_2, \nn \\
&+& T^2 \, B_T  + \frac{3 \, T^4 \, \lambda_2}{4 \, f^2}. 
\end{eqnarray}

\subsection{Thermal masses: Vector Bosons}

When employing ring resummation in the SM, a very important  effect is the limit the magnetic mass places on the Higgs self coupling for a first order EWPT to occur.  The magnetic mass  is a 
nonperturbative contribution to the transverse mass in thermal field theory, the inverse of which corresponds to the magnetic screening length for the $SU(2)$ sector of theory. Although
we cannot calculate the magnetic mass, we can estimate its effects in perturbation theory
when one calculates the gauge polarization tensor in one loop gap equations.
The magnetic mass  still imposes a very important constraint on the phase transition even in our HEFT. As we are examining deviations from the SM in PGH models, it is appropriate to have the SM calculation of the gauge polarization tensor in thermal field theory performed as accurately as possible. Thus we determine the one loop gap equations for the gauge boson degrees of freedom.
The requisite diagrams to calculate are given in Fig. 12 when the propagators are full propagators
whose masses are dictated by the self consistent solution of the one loop gap equations.

\vspace{1cm}
\begin{figure} \label{vectors}
\centerline{\scalebox{0.7}{\includegraphics{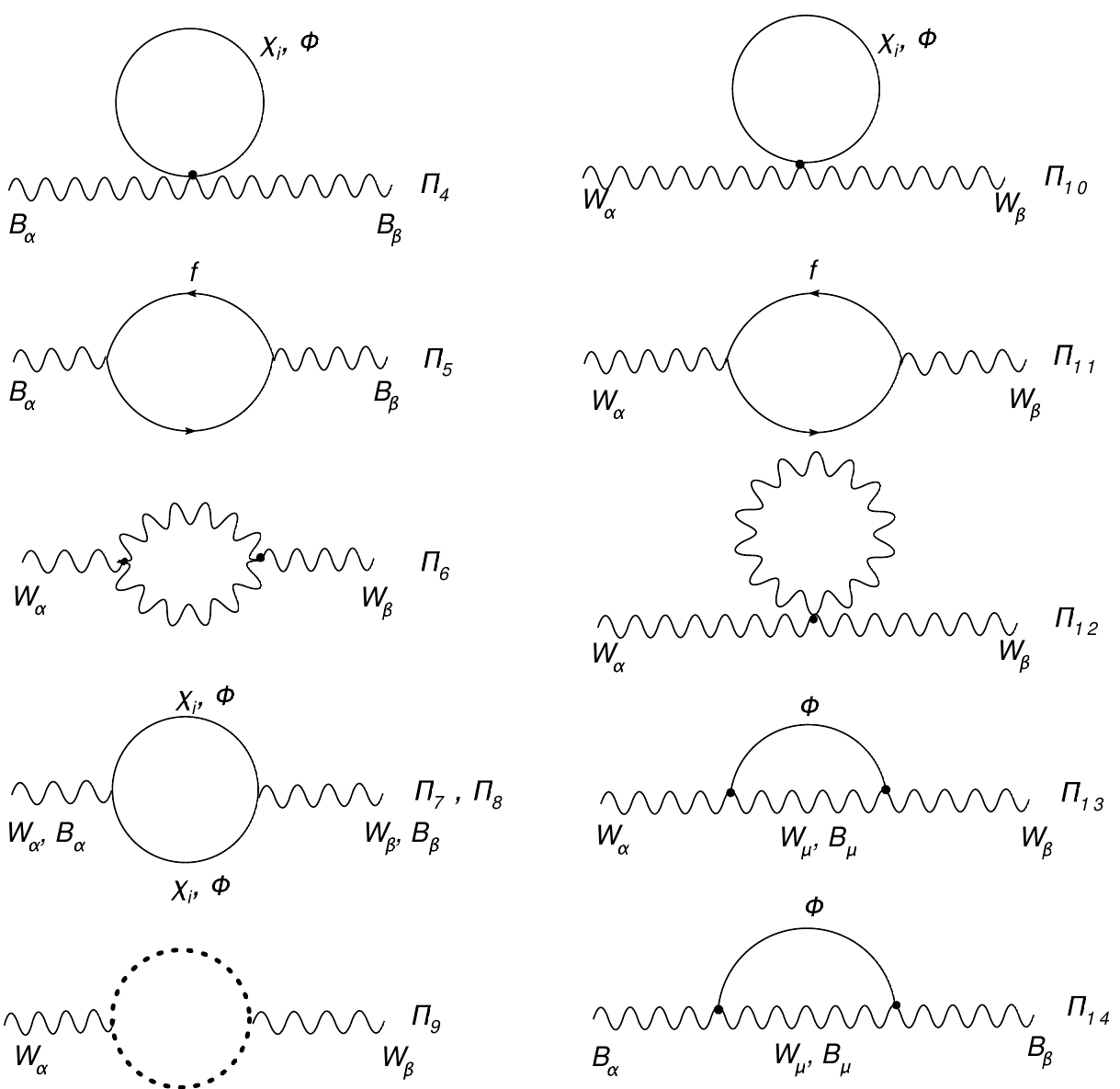}}}
\caption{One loop diagrams that contribute to the vector boson thermal masses. $f$ indicates a sum over all spin $1/2$ particles.}
\end{figure}

Again we seek to  obtain thermal mass contributions appropriate to shift the mass in ring resummation and set the external momenta $(k_0, \bf{k})$
to zero by setting $k_0 = 0$ and taking the limit ${\bf k^2} \rightarrow 0$. 
The complications involved in considering mixing originate from the asymmetry
between the temporal and spatial components in thermal field theory. Due to this asymmetry the 
longitudinal (temporal) and transverse (spatial) modes of the gauge field develop different effective masses at finite temperature. Thus we decompose the propagator in Landau gauge\footnote{As this decomposition is gauge dependent the resultant thermal masses  and thermal Weinberg angles will be gauge dependent. In fact the effective potential itself is gauge dependent as well, however,
all physical quantities derived from the effective potential will be gauge independent. Note there exists a subtlety in the decomposition that leads to a factor of 2/3\cite{Buchmuller:1993bq} .} as 
\begin{eqnarray}
i D_i^{\mu \, \nu}(k) = \frac{P_L^{\mu \, \nu}}{k^2 - m_i^2 - \Pi_L(k)} + \frac{P_T^{\mu \, \nu}}{k^2 - m_i^2 - \Pi_T(k)}, 
\end{eqnarray}
where the transverse and longitudinal projectors are
\begin{eqnarray}
P_T^{\mu \, \nu} &=& g^{\mu}_i \left(\delta^{i \,  j} - \frac{k^i \, k^j}{\bf{k}^2} \right) \, g^{\nu}_j, \\
P_L^{\mu \, \nu} &=&  \frac{k^\mu \, k^\nu}{k^2} - g^{\mu \, \nu} - P_T^{\mu \, \nu}, 
\end{eqnarray}
and $m_i^2$ is the tree level mass.  One can determine the transverse and longitudinal 
corrections to the mass via $\Pi_L(0) = -\Pi^0_0$ and $\Pi_T(0) = -\Pi^{i}_i(0)/3$. The difficulty is that 
once a mass eigenstate basis is  known then the corrections to the massive gauge bosons are easy to determine,
however what exactly the mass eigenstate basis is depends on thermal corrections. We circumvent this difficulty by
first calculating the diagrams with no internal massive gauge bosons in the $B_\mu, W^I_\mu$ basis.  
For the longitudinal mass the results of the diagrams $\Pi_4, \Pi_5, \Pi_7, \Pi_8, \Pi_9, \Pi_{10}, \Pi_{11}$ are 
\begin{eqnarray}
\langle W^a \, W^b \rangle_L^1 &=& (\Pi_7 + \Pi_9 +  \Pi_{10}+ \Pi_{11})_L, \nn \\
&=& - \frac{g_2^2 \, T^2 \, \delta^{ab}}{\pi^2} \left[F_3(y_\chi^2, y_h^2) + \frac{I_+(0)}{2} \right], \nn \\
&+& \frac{g_2^2 \, T^2 \, \delta^{ab}}{\pi^2} \left[\frac{I_+(y_h^2) + 3 I_+(y_\chi^2)}{8}\right], \\
&-& \frac{12 \, g_2^2 \, T^2 \, \delta^{ab}}{\pi^2} \left[I_-(y_t^2) - y_t^2 K_-(y_t^2)/2\right], \nn
\end{eqnarray}
\begin{eqnarray}
\langle B B \rangle_L^1  &=& (\Pi_4 + \Pi_5 + \Pi_8)_L, \nn \\
&=& - \frac{g_1^2 \, T^2 \, \delta^{ab}}{\pi^2} \left[F_3(y_\chi^2, y_h^2) \right], \nn \\
&+& \frac{g_1^2 \, T^2 \, \delta^{ab}}{\pi^2} \left[\frac{I_+(y_h^2) + 3 I_+(y_\chi^2)}{8}\right], \\
&-& \frac{20 \, g_1^2 \, T^2 \, \delta^{ab}}{\pi^2} \left[I_-(y_t^2) - y_t^2 K_-(y_t^2)/2\right], \nn \\
\langle W^3 \, B \rangle_L^1 &=&  \frac{g_1 \, g_2 \, T^2}{8 \, \pi^2} \left[ I_+(y_\chi^2) - I_+(y_h^2) \right].
\end{eqnarray}
We then rotate these contributions to the two point functions by assuming that an angle exists  for any $T$ to diagonalize the external $W_3, B$ fields. This defines a thermal basis of the bosonic fields with 
\begin{eqnarray}
Z &=& \cos(\theta(T)) \, W^3 -  \sin(\theta(T)) \, B, \\
A &=& \sin(\theta(T)) \, W^3 +  \cos(\theta(T)) \, B,  \nn
\end{eqnarray}
and $W^{\pm}$ is  related to $W^{1,2}$ in the usual manner. This thermal angle will limit to the Weinberg
angle as $T \rightarrow 0$.  The remaining contributions to the projected two point functions 
are obtained from 
the diagrams $\Pi_6, \Pi_{12}, \Pi_{13}, \Pi_{14}$. One finds 
\begin{eqnarray}
\langle W^+ \, W^- \rangle_L^2 &=&  - \frac{8 \, g_2^2 \sin^2(\theta(T)) \, T^2}{\pi^2} \,  F_1[(y_W^T)^2,(y_A^T)^2], \nn \\ 
&-&  \frac{4 \, g_2^2 \sin^2(\theta(T)) \, T^2}{\pi^2} \,  F_1[(y_W^L)^2,(y_A^L)^2],  \nn\\
&-& \frac{8 \, g_2^2 \cos^2(\theta(T)) \, T^2}{\pi^2} \,  F_1[(y_W^T)^2,(y_Z^T)^2], \nn \\ 
&-&  \frac{4 \, g_2^2 \cos^2(\theta(T)) \, T^2}{\pi^2} \,  F_1[(y_W^L)^2,(y_Z^L)^2],  \nn\\
&+& \frac{g_2^2 \, T^2 \, \sin^2(\theta(T))}{2 \, \pi^2} \, F_4[(y_A^L)^2,(y_A^T)^2] ,   \\
&+&  \frac{g_2^2 \, T^2 \, \cos^2(\theta(T))}{2 \, \pi^2} \, F_4[(y_Z^L)^2,(y_Z^T)^2],  \nn  \\
&+&  \frac{g_2^2 \, T^2}{2 \, \pi^2} \, F_4[(y_W^L)^2,(y_W^T)^2], \nn \\
&+& \frac{g_2^4}{4} \,  F_5[(y_W^L)^2,(y_{\phi})^2], \nn \\
&+& \frac{g_1^2 \, g_2^2}{4} \, \cos^2(\theta(T)) \,  F_5[(y_A^L)^2,(y_{\chi})^2], \nn \\
&+& \frac{g_1^2 \, g_2^2 }{4} \, \sin^2(\theta(T)) \,  F_5[(y_Z^L)^2,(y_{\chi})^2]. \nn
\end{eqnarray}
The high temperature expansion of $(m_W^L(T,\varphi))^2$ is given by
\begin{eqnarray}
(m_W^L(T,\varphi))^2&=&  \frac{11 g_2^2 \, T^2}{6} + \frac{g_2^2 \, \varphi^2}{4} - \frac{g_2^4 \varphi^2 \, T}{16 \, \pi \left(m_h + m_W^L \right)}, \nn \\
&-&\frac{g_2^2 \, T}{16 \, \pi} \left(m_h + 3 \, m_\chi + 4 m_Z^L + 8 m_W^L + 4 m_A^L \right), \nn \\
&+& \frac{g_2^2 \, T(m_A^L - m_Z^L) \, \cos(2\, \theta(T))}{4 \, \pi} \\
&-& \frac{g_1^2 \, g_2^2 \, v^2 \, T}{16 \, \pi} \left(\frac{\sin^2(\theta(T))}{m_Z^L + m_\chi} + \frac{\cos^2(\theta(T))}{m_A^L + m_\chi} \right). \nn
\end{eqnarray}
which reproduces the known answer for the case of vanishing $U(1)$ charge  \cite{Buchmuller:1993bq} in the $g_1 \rightarrow 0$, $\theta \rightarrow 0$ limit. Note that we have added the 
usual EW term $g_2^2 \, \varphi^2/4$ to this expression.
For thermal photon and $Z$ fields  one finds the following
\begin{eqnarray}
\langle A \, A \rangle_L^2 &=&   - \frac{8 \, g_2^2 \sin^2(\theta(T)) \, T^2}{\pi^2} \,  F_2[(y_W^T)^2], \nn \\ 
&-&  \frac{4 \, g_2^2 \sin^2(\theta(T)) \, T^2}{\pi^2} \,  F_2[(y_W^L)^2],  \nn\\
&+& \frac{g_2^2 \, T^2 \, \sin^2(\theta(T))}{\pi^2} \, F_4[(y_W^L)^2,(y_W^T)^2] ,   \\
&+& \frac{g_1^2 \, g_2^2}{4} \, \cos^2(\theta(T)) \,  F_5[(y_W^L)^2,(y_{\chi})^2], \nn \\
&+& \frac{(g_1 \, \cos(\theta(T)) - g_2 \, \sin(\theta(T)))^4}{8} \, F_5[(y_A^L)^2,(y_{h})^2] \nn
\end{eqnarray}
\begin{eqnarray}
\langle Z \, Z \rangle_L^2 &=&   - \frac{8 \, g_2^2 \cos^2(\theta(T)) \, T^2}{\pi^2} \,  F_2[(y_W^T)^2], \nn \\ 
&-&  \frac{4 \, g_2^2 \cos^2(\theta(T)) \, T^2}{\pi^2} \,  F_2[(y_W^L)^2],  \nn\\
&+& \frac{g_2^2 \, T^2 \, \cos^2(\theta(T))}{\pi^2} \, F_4[(y_W^L)^2,(y_W^T)^2] ,   \\
&+& \frac{g_1^2 \, g_2^2}{4} \, \sin^2(\theta(T)) \,  F_5[(y_W^L)^2,(y_{\chi})^2], \nn \\
&+& \frac{(g_1 \, \cos(\theta(T)) + g_2 \, \sin(\theta(T)))^4}{8} \, F_5[(y_Z^L)^2,(y_{h})^2]. \nn
\end{eqnarray}
\begin{eqnarray}
\langle A \, Z \rangle_L^2 &=&   - \frac{8 \, g_2^2 \cos(\theta(T)) \,  \sin(\theta(T)) \, T^2}{\pi^2} \,  F_2[(y_W^T)^2], \nn \\ 
&-&  \frac{4 \, g_2^2 \cos(\theta(T)) \,  \sin(\theta(T)) \, T^2}{\pi^2} \,  F_2[(y_W^T)^2],   \\
&+& \frac{g_2^2 \, T^2 \, \sin(\theta(T)) \, \cos(\theta(T))}{\pi^2} \, F_4[(y_W^L)^2,(y_W^T)^2] ,   \nn \\
&+& \frac{g_1^2 \, g_2^2}{4} \, \cos(\theta(T)) \, \sin(\theta(T))  \,  F_5[(y_W^L)^2,(y_{\chi})^2], \nn
\end{eqnarray}
The thermal Weinberg angle $\theta(T)$ for the longitudinal mass is defined by demanding that 
\begin{eqnarray}
(m_{AZ}^L)^2  &=&  \sin(\theta(T)) \, \cos(\theta(T))(\langle W^3 \, W^3 \rangle_L^1 -  \langle B \, B \rangle_L^1), \nn \\ 
&+& (\cos^2(\theta(T))  - \sin^2(\theta(T)))  \langle W^3 \, B \rangle_L^1 +  \langle A \, Z \rangle_L^2 \nn
\end{eqnarray}
vanish for a specific T. These expressions are rather daunting. Let us first consider the case of vanishingly small temperature.
Adding the usual tree level EW terms the expression $(m_{AZ}^L)^2$ reduces to 
\begin{eqnarray}
(m_{AZ}^L)_{T \rightarrow 0}^2 &=& - \frac{\varphi^2}{8} \, \left(2 \, g_1 \, g_2 \cos(2 \theta(T))\right) \nn \\
&-&  \frac{\varphi^2}{8} \, \left(( g_1^2 - g_2^2) \sin(2 \, \theta(T)) \right),
\end{eqnarray}
demanding that $(m_{AZ}^L)^2 = 0$ the solution is 
\begin{eqnarray}
\sin(\theta(0)) = \frac{g_1}{\sqrt{g_1^2 + g_2^2}},
\end{eqnarray}
which establishes that in the limit $T \rightarrow 0$ thermal Weinberg angle reduces to the usual $\theta_W$.
In the opposite limit as $T \gg \varphi $ one finds the leading term
\begin{eqnarray}
(m_{AZ}^L)_{\varphi \rightarrow 0}^2&=& \frac{11 \,  T^2}{12} \, \left(g_2^2 - g_1^2 \right) \,  \sin(2 \, \theta(T)),
\end{eqnarray}
and $(m_{AZ}^L)^2$  for $\sin(\theta(T)) = 0$ at high temperature. As expected, EW symmetry is 
restored and the diagonal basis is the basis of the unbroken electroweak theory $W^I_\mu, B_\mu$. 
Although not unexpected, this is entertaining.

Employing the high temperature expansion we can perturbatively solve as the temperature
decreases. The thermal angle will be $\sin(\theta(T)) = 0 + \mathcal{O}(1/T)$ and we find that the temperature dependence of
thermal Weinberg angle is 
\begin{eqnarray}
\sin(\theta(T))_{T>\varphi} = \frac{3 \, g_1 \, g_2 (m_h(\varphi,T) - m_\chi(\varphi,T))}{88 \, ( g_2^2 - g_1^2) \, \pi \, T}
\end{eqnarray}
so that as the temperature lowers, thermal Weinberg angle rises toward $\theta_W$ 
and the correct basis changes over to the basis in the broken electroweak theory. Note that the scalar masses are
the results of the one loop gap equation scalar masses that are not imaginary for small $\varphi$.

However, we are interested in mass effects when the temperature is eventually approaching the temperature $T_b$ not this extreme case.
We retain terms of $\mathcal{O}(g_{sm}^2 \varphi^2)$ and  $\mathcal{O}(g_{sm}^2 T^2)$ when solving the equation.  Note that
terms of order $\mathcal{O}(g_{sm}^2 \varphi^2)$ from the high temperature expansion are loop suppressed compared to the tree level 
electroweak terms and are dropped. The temperature scale is set 
by $T_b \sim 100 \, {\rm GeV}$ for the phase transition and $\varphi \lesssim T_b < v$ as the minima is not yet reached in the potential.
Thus, one can see that the high temperature expansion is properly thought of as a perturbative expansion in $g_{sm}$ for the 
temperatures of interest about the phase transition. We solve $(m_{AZ}^L)^2$ perturbatively in $g_{SM}$ for this reason.
The resulting expression is still rather daunting.  However the physical dependence can be deduced by using $g_2^2 = 4 M_W^2(v)/v^2$
and  $g_1^2 = 4 (M_Z^2(v) - M_W^2(v))/v^2$ and expanding in $(T-\varphi)$. One finds
\begin{eqnarray}
\sin(\theta(T)) \approx 0.09 - \frac{0.15(T-\varphi)}{T} + \frac{0.03 (T-\varphi)^2}{T^2}. \nn 
\end{eqnarray}
With these insights we can state the correct longitudinal thermal masses for the bosonic fields relevant
for studies of the EWPT to be, to $\mathcal{O}(g_{SM}^2)$
\begin{eqnarray}
(m^L_W(\varphi,T))^2 &=&  g_2^2 \, \left(\frac{11 \, T^2}{6} + \frac{\varphi^2}{4} \right), \\
(m^L_A(\varphi,T))^2 &=&   \frac{11  T^2}{6}  \left(g_1^2 \,  \cos^2(\theta(T)) + g_2^2\,  \sin^2(\theta(T))  \right) \nn \\
&\,& + \frac{\varphi^2}{4} \left(g_1 \, \cos(\theta(T)) - g_2 \, \sin(\theta(T))\right)^2, \nn \\
(m^L_Z(\varphi,T))^2 &=&  \frac{11 T^2}{6} \left(g_2^2 \,  \cos^2(\theta(T)) + g_1^2\,  \sin^2(\theta(T))  \right) \nn \\
&\,& + \frac{\varphi^2}{4} \left(g_1 \, \sin(\theta(T)) +g_2 \, \cos(\theta(T))\right)^2. \nn
\end{eqnarray}
The appropriate approximation for thermal Weinberg angle in studies of the electroweak phase transition
is $\sin(\theta(T_b))$. However,  our formalism can be used for 
numerical studies not using this approximation if desired.

The calculations for the transverse mass are similar. For the transverse mass the results for the diagrams
$\Pi_4, \Pi_5, \Pi_7, \Pi_8, \Pi_9, \Pi_{10}, \Pi_{11}$ are 
\begin{eqnarray}
\langle W^a \, W^b \rangle_T^1 &=& (\Pi_7 + \Pi_9 +  \Pi_{10}+ \Pi_{11})_L, \nn \\
&=& - \frac{g_2^2 \, T^2 \, \delta^{ab}}{4 \, \pi^2} \left[I_+(y_\chi^2) + \frac{2\,  \left(J_+(y_\chi^2) - J_+(y_h^2) \right)}{y_\chi^2 - y_h^2}\right], \nn \\
&+& \frac{g_2^2 \, T^2 \, \delta^{ab}}{\pi^2} \left[\frac{I_+(y_h^2) + 3 I_+(y_\chi^2)}{8}\right], \\
&+& \frac{6 \, g_2^2 \, T^2 \, \delta^{ab}}{\pi^2} \left[y_t^2 K_-(y_t^2)\right] + \frac{g_2^2 \, T^2 \, \delta^{ab}}{2 \pi^2} \, I_+(0), \nn 
\end{eqnarray}
\begin{eqnarray}
\langle B B \rangle_T^1  &=& (\Pi_4 + \Pi_5 + \Pi_8)_T, \nn \\
&=& - \frac{g_1^2 \, T^2 \, \delta^{ab}}{4 \, \pi^2} \left[I_+(y_\chi^2) + \frac{2\,  \left(J_+(y_\chi^2) - J_+(y_h^2) \right)}{y_\chi^2 - y_h^2}\right], \nn \\
&+& \frac{g_1^2 \, T^2 \, \delta^{ab}}{\pi^2} \left[\frac{I_+(y_h^2) + 3 I_+(y_\chi^2)}{8}\right], \\
&+& \frac{20 \, g_1^2 \, T^2 \, \delta^{ab}}{\pi^2} \left[y_t^2 \,  K_-(y_t^2)\right], \nn \\
\langle W^3 \, B \rangle_T^1 &=&  \frac{g_1 \, g_2 \, T^2}{8 \, \pi^2} \left[ I_+(y_\chi^2) - I_+(y_h^2) \right].
\end{eqnarray}

Again we rotate these contributions to the two point functions by assuming that an angle exists to
diagonalize the external $W_3, B$ fields. Note however that this angle is not the same as in the 
longitudinal case although it remains true that this second  thermal angle will limit to the Weinberg
angle as $T \rightarrow 0$. Again this defines a thermal basis of fields  with 
\begin{eqnarray}
Z &=& \cos(\theta'(T)) \, W^3 -  \sin(\theta'(T)) \, B, \\
A &=& \sin(\theta'(T)) \, W^3 +  \cos(\theta'(T)) \, B,  \nn
\end{eqnarray}
and $W^{\pm}$ related to $W^{1,2}$ in the usual manner. For
the diagrams $\Pi_6, \Pi_{12}, \Pi_{13}, \Pi_{14}$,  one finds for the transverse mass
\begin{eqnarray}
\langle W^+ \, W^- \rangle_T^2 &=& \frac{8 \, g_2^2 \sin^2(\theta'(T)) \, T^2}{3 \pi^2} \,  F_6[(y_W^T)^2,(y_A^T)^2], \nn \\ 
&+&  \frac{4 \, g_2^2 \sin^2(\theta'(T)) \, T^2}{3 \, \pi^2} \,  F_6[(y_W^L)^2,(y_A^L)^2],  \nn\\
&+& \frac{8 \, g_2^2 \cos^2(\theta'(T)) \, T^2}{3 \, \pi^2} \,  F_6[(y_W^T)^2,(y_Z^T)^2], \nn \\ 
&+&  \frac{4g_2^2 \cos^2(\theta'(T)) \, T^2}{3 \, \pi^2} \,  F_6[(y_W^L)^2,(y_Z^L)^2],  \nn\\
&+& \frac{g_2^2 \, T^2 \, \sin^2(\theta'(T))}{3 \, \pi^2} \, F_7[(y_A^L)^2,(y_A^T)^2] ,   \\
&+&  \frac{g_2^2 \, T^2 \, \cos^2(\theta'(T))}{3 \, \pi^2} \, F_7[(y_Z^L)^2,(y_Z^T)^2],  \nn  \\
&+&  \frac{g_2^2 \, T^2}{3 \, \pi^2} \, F_7[(y_W^L)^2,(y_W^T)^2] + \frac{g_2^4}{4} \,  F_8[(y_W^L)^2,(y_{\phi})^2], \nn \\
&+& \frac{g_1^2 \, g_2^2}{4} \, \cos^2(\theta'(T)) \,  F_8[(y_A^L)^2,(y_{\chi})^2], \nn \\
&+& \frac{g_1^2 \, g_2^2 }{4} \, \sin^2(\theta'(T)) \,  F_8[(y_Z^L)^2,(y_{\chi})^2]. \nn
\end{eqnarray}
So that the high temperature expansion of $\langle W^+ \, W^- \rangle_T^1 + \langle W^+ \, W^- \rangle_T^2$ is given by
\begin{eqnarray}
(m_{W}^T)^2&=& \frac{g_2^2 \, m_W^T \, T}{3 \, \pi} + \frac{g_2^2 \, m_W^L \, T}{12 \, \pi}  + \frac{g_2^2 \, \varphi^2}{4}, \nn \\
&+& \frac{g_2^2 \,T \, 5 \, \left(m_Z^T \, \cos^2[\theta'(T)]  + m_A^T \, \sin^2[\theta'(T)] \right)}{12 \, \pi} \nn \\
&-& \frac{g_1^2 \, g_2^2 \, \varphi^2 \, T \, \cos^2[\theta'(T)] }{24 \, \pi (m_A^L + m_\chi)} - \frac{g_1^2 \, g_2^2 \, \varphi^2 \, T \, \sin^2[\theta'(T)] }{24 \, \pi (m_Z^L + m_\chi)} \nn \\
&-& \frac{2 \, g_2^2 \, m_W^T \, m_Z^T \, T \, \cos^2[\theta'(T)] }{3 \, \pi (m_W^T + m_Z^T)} + \frac{g_2^2 \, T \, (m_h - m_\chi)^2}{48 \, \pi \, (m_h + m_\chi)} \nn \\
&-& \frac{2 \, g_2^2 \, m_W^T \, m_A^T \, T \, \sin^2[\theta'(T)] }{3 \, \pi (m_W^T + m_A^T)} - \frac{g_2^4 \, \varphi^2 \, T}{24 \, \pi \,  (m_h + m_W^L)} \nn \\
&-& \frac{g_2^2 \, m_W^L \, m_Z^L \, T \, \cos^2[\theta'(T)] }{3 \, \pi (m_W^L + m_Z^L)}  \\
&-& \frac{g_2^2 \, m_W^L \, m_A^L \, T \, \sin^2[\theta'(T)] }{3 \, \pi (m_W^L + m_A^L)} \nn
\end{eqnarray}
which also reproduces the known answer  for the case of vanishing $U(1)$ charge  \cite{Buchmuller:1993bq}  in the $g_1 \rightarrow 0$, $\theta \rightarrow 0$ limit. The transverse masses of thermal photon and $Z$ fields are deduced from 
the addition of the rotated contributions and the following 
\begin{eqnarray}
\langle A \, A \rangle_T^2 &=&   - \frac{g_2^2 \sin^2(\theta'(T)) \, T^2}{\pi^2} \,  \left(I_+[(y_W^L)^2] + 2 \, I_+[(y_W^T)^2] \right), \nn \\ 
&+& \frac{2 \, g_2^2 \, T^2 \, \sin^2(\theta'(T))}{3 \, \pi^2} \, F_7[(y_W^L)^2,(y_W^T)^2] ,   \\
&+& \frac{g_1^2 \, g_2^2}{4} \, \cos^2(\theta(T)) \,  F_8[(y_W^L)^2,(y_{\chi})^2], \nn \\
&+& \frac{(g_1 \, \cos(\theta(T)) - g_2 \, \sin(\theta(T)))^4}{8} \, F_8[(y_A^L)^2,(y_{h})^2] \nn
\end{eqnarray}
\begin{eqnarray}
\langle Z \, Z \rangle_T^2 &=&   - \frac{g_2^2 \cos^2(\theta'(T)) \, T^2}{\pi^2} \,  \left(I_+[(y_W^L)^2] + 2 \, I_+[(y_W^T)^2] \right), \nn \\ 
&+& \frac{2 \, g_2^2 \, T^2 \, \cos^2(\theta'(T))}{3 \, \pi^2} \, F_7[(y_W^L)^2,(y_W^T)^2] ,   \\
&+& \frac{g_1^2 \, g_2^2}{4} \, \sin^2(\theta(T)) \,  F_8[(y_W^L)^2,(y_{\chi})^2], \nn \\
&+& \frac{(g_1 \, \cos(\theta(T)) + g_2 \, \sin(\theta(T)))^4}{8} \, F_8[(y_Z^L)^2,(y_{h})^2] \nn
\end{eqnarray}
\begin{eqnarray}
\langle A \, Z \rangle_T^2 &=&   - \frac{g_2^2 \cos(\theta'(T)) \, \sin(\theta'(T)) \, T^2}{\pi^2} \, I_+[(y_W^L)^2] \nn \\
&-&  \frac{2 \, g_2^2 \cos(\theta'(T)) \, \sin(\theta'(T)) \, T^2}{\pi^2} \,  I_+[(y_W^T)^2], \nn \\ 
&+& \frac{2 \, g_2^2 \, T^2 \, \sin(\theta'(T)) \, \cos(\theta'(T))}{3 \,\pi^2} \, F_7[(y_W^L)^2,(y_W^T)^2] ,   \nn \\
&+& \frac{g_1^2 \, g_2^2}{4} \, \cos(\theta'(T)) \, \sin(\theta'(T))  \,  F_8[(y_W^L)^2,(y_{\chi})^2], \nn
\end{eqnarray}
As in the longitudinal case, thermal Weinberg angle $\theta'(T)$ for the transverse mass is defined by demanding that 
\begin{eqnarray}
(m_{AZ}^T)^2 &=&  \sin(\theta'(T)) \, \cos(\theta'(T))(\langle W^3 \, W^3 \rangle_T^1 -  \langle B \, B \rangle_T^1), \nn \\ 
&+& (\cos^2(\theta'(T))  - \sin^2(\theta'(T)))  \langle W^3 \, B \rangle_T^1 +  \langle A \, Z \rangle_T^2 \nn
\end{eqnarray}
vanish for a given T. Again at small temperature we find 
\begin{eqnarray}
(m_{AZ}^T)^2_{T \rightarrow 0} &=& - \frac{\varphi^2}{8} \, \left(2 \, g_1 \, g_2 \cos(2 \theta'(T))\right) \nn \\
&-&  \frac{\varphi^2}{8} \, \left(( g_1^2 - g_2^2) \sin(2 \, \theta'(T)) \right),
\end{eqnarray}
however at high temperature we now have 
\begin{eqnarray}
(m_{AZ}^T)^2_{\varphi \rightarrow 0} &=& - \frac{T^2}{48} \,  g_1^2 \, \sin(2 \, \theta'(T)),
\end{eqnarray}
however, we again have that  $(m_{AZ}^T)^2 = 0$  for $\sin(\theta'(T)) = 0$ at high temperature.
Numerically approximating the solution as before and expanding in $(T-\varphi)$. One finds
\begin{eqnarray}
\sin(\theta'(T)) \approx 0.49 + \frac{0.03(T-\varphi)}{T} + \frac{0.04(T-\varphi)^2}{T^2}. \nn 
\end{eqnarray}
from which we see that at $\varphi \lesssim T_b$ the transverse masses have already mixed far further into the thermal mass basis
 from the initial EW basis.  
The transverse thermal masses for the bosonic fields relevant
for studies of the EWPT to $\mathcal{O}(g_{SM}^3)$ are
\begin{eqnarray}\label{transeqn}
(m^T_W(\varphi,T))^2 &=& (m_{W}^T)^2(\varphi,T), \\
(m^T_A(\varphi,T))^2 &=&   \frac{g_2^2 \, m_W^T \, T \, \sin^2[\theta'(T)]}{3 \, \pi} + \frac{g_1^2 \, T^2 \, \cos^2[\theta'(T)]}{24}, \nn \\
&+&  \frac{\varphi^2 \left(g_2 \, \sin[\theta'(T)] - g_1 \, \cos[\theta'(T)] \right)^2}{4}, \nn \\
&+& F_+(m_h(\varphi),m_\chi(\varphi) ), \nn \\
(m^T_Z(\varphi,T))^2 &=& \frac{g_2^2 \, m_W^T \, T \, \cos^2[\theta'(T)]}{3 \, \pi} + \frac{g_1^2 \, T^2 \, \sin^2[\theta'(T)]}{24}, \nn \\
&+&  \frac{\varphi^2 \left(g_2 \, \cos[\theta'(T)] + g_1 \, \sin[\theta'(T)] \right)^2}{4}, \nn \\
&+& F_-(m_h(\varphi),m_\chi(\varphi) ).
\end{eqnarray}
We have defined the following functions of the scalar masses
\begin{eqnarray}
F_{\pm}(m_h(\varphi),m_\chi(\varphi) ) &=& A^1_\varphi \left(g_1^2 + g_2^2\right), \nn \\
&\pm&   A^1_\varphi \left(g_1^2 - g_2^2\right)  \, \cos[2 \, \theta'(T)] \nn \\
&\pm&   6 \, A^2_\varphi \, g_1 \, g_2  \, \sin[2 \, \theta'(T)],
\end{eqnarray}
\begin{eqnarray}
A_\varphi^1 &=& \frac{T}{96 \, \pi} \frac{(m_h(\varphi) - m_\chi(\varphi))}{(m_h(\varphi) +m_\chi(\varphi))} (m_h(\varphi) - m_\chi(\varphi)), \nn \\
A_\varphi^2 &=& \frac{T}{96 \, \pi} \frac{(m_h(\varphi) - m_\chi(\varphi))}{(m_h(\varphi) +m_\chi(\varphi))} (m_h(\varphi) + m_\chi(\varphi)). \nn 
\end{eqnarray}
We have not in fact solved for the transverse masses as yet due to the appearance of $m_W^T$
on both sides of Eqn. (\ref{transeqn}).  This is an important feature of the transverse mass that leads to the 
inclusion of a nonperturbative magnetic mass term. Consider $\varphi \rightarrow 0$, then $\cos({\theta'}(T)) = 1, \sin(\theta'(T)) = 0$ and $m_h(0) = m_\chi(0)$. In this case, the expression for $m_W^T$ is given by
\begin{eqnarray}
m^T_W(0,T)^2 = \frac{g_2^2 \, T}{3 \, \pi} \, m^T_W(0,T). 
\end{eqnarray}
The  physical solution  \cite{Espinosa:1992kf} is 
\begin{eqnarray}
m^T_W(0,T) = \frac{g_2^2 \, T}{3 \, \pi}.
\end{eqnarray}
A non-abelian gauge theory is expected to have a term of this form
as a nonperturbative feature \cite{Linde:1980ts,Gross:1980br}. Of course, we cannot calculate a nonperturbative
result in perturbation theory.
We retain this term as it plays an important role as $\varphi \rightarrow 0$ in determining the nature of the phase transition. We multiply the magnetic mass term by an unknown $\gamma$
factor to signify its nonperturbative origin in our $V_{eff}^{ring}$.
Lattice simulations have determined $m_m(T) = 0.456(6) \, g_2^2(T) \, T$  \cite{Heller:1997nqa} which one expects to be a 
good approximation of the magnetic mass of the $\rm SU(2)$ sector of the SM, giving $\gamma= 4.2$.

The magnetic mass for $m^T_W(0,T)$
contributes a magnetic mass term to $m^T_Z(0,T)$ and $m^T_A(0,T)$. 
For $m^T_A(0,T)$ there is a $g_1$ term that is $\varphi$ independent.
As $\varphi \rightarrow 0$ the magnetic mass does not screen thermal photon field.
Thus the magnetic mass effects on the photon field can be dropped.

For $m^T_Z(\varphi,T)$, in the  $\varphi \rightarrow 0$ limit the magnetic mass does screen the 
$Z$ field and is retained. The $\mathcal{O}(g_{SM}^2)$ transverse masses, including only the important $\varphi$ independent magnetic mass terms that are higher order are given by
\begin{eqnarray}
(m^T_W(\varphi,T))^2 &=&\frac{\gamma^2 \, g_2^4}{9 \, \pi^2} \, T^2 + \frac{g_2^2 \, \varphi^2}{4}, \\
(m^T_A(\varphi,T))^2 &=&   \frac{g_1^2 \, T^2 \, \cos^2[\theta'(T)]}{24}, \nn \\
&+&  \frac{\varphi^2 \left(g_2 \, \sin[\theta'(T)] - g_1 \, \cos[\theta'(T)] \right)^2}{4}. \nn \\
(m^T_Z(\varphi,T))^2 &=& \frac{g_2^2 \, \, m^T_W(\varphi,T) \, T \, \cos^2[\theta'(T)]}{3 \, \pi} \nn \\
&+&  \frac{g_1^2 \, T^2 \, \sin^2[\theta'(T)]}{24}, \nn \\
&+&  \frac{\varphi^2 \left(g_2 \, \cos[\theta'(T)] + g_1 \, \sin[\theta'(T)] \right)^2}{4}. \nn
\end{eqnarray}
where we have neglected the functions $F_{\pm}$ that are $\mathcal{O}(g_{SM}^2 \, \lambda_1)$
and suppressed by loop factors.

\subsection{$\mathcal{O}(\phi^\dagger \, \phi)^3$ Finite Temperature Terms}

As we are considering the effects of the operator $(\phi^\dagger \, \phi)^3$ one 
should note that matching corrections to this operator are obtained by expanding the one loop 
finite temperature contributions given in Eqn.~(\ref{finitet}).

Expanding Eqn.~(\ref{finitet}) to higher order it is easy to retain the $m^6$ term. 
For the bosons this gives a contribution 
\begin{eqnarray}
V^B_6(T) = \frac{(\phi^\dagger \, \phi)^3}{\Lambda^2} \frac{\zeta(3) \, \Lambda^2}{6 \, (4 \, \pi^2)^2 \, T^2} \, \left( \frac{m_h^6}{v^6} + 6 \, \frac{m_W^6}{v^6} + 3 \, \frac{m_Z^6}{v^6} \right),\nn
\end{eqnarray}
to the potential. For the fermions we similarly have 
\begin{eqnarray}
V^F_6(T) = \frac{(\phi^\dagger \, \phi)^3}{\Lambda^2} \, \frac{8 \, \left(7 \, \zeta(3) - 8\right) \, \Lambda^2}{(4 \, \pi^2)^2 \, T^2} \, \frac{m_t^6}{v^6}
\end{eqnarray}
These matching corrections are small for the temperatures of interest.
For $m_h = 120 \, {\rm GeV},v = 246 \, {\rm Gev}$ and the PGD values \cite{PDBook} for the known masses 
$m_W = 80.4 \, {\rm GeV},m_Z = 91.2 \, {\rm GeV},m_t = 172.5 \, {\rm GeV}$ the sum of the matching corrections
above gives a correction of size $
2.6 \times 10^{-4}/T^2, $
to the $(\phi^\dagger \, \phi)^3$ operator. Due to its negligible coefficient we neglect this matching correction. 
\vfill
\bibliographystyle{h-physrev3.bst}
\bibliography{Higgs}

\begin{thebibliography}{10}

\bibitem{Barate:2003sz}
LEP Working Group for Higgs boson searches, R.~Barate {\em et~al.},
\newblock Phys. Lett. {\bf B565}, 61 (2003), hep-ex/0306033.

\bibitem{Georgi:1975tz}
H.~Georgi and A.~Pais,
\newblock Phys. Rev. {\bf D12}, 508 (1975).

\bibitem{Kaplan:1983sm}
D.~B. Kaplan, H.~Georgi, and S.~Dimopoulos,
\newblock Phys. Lett. {\bf B136}, 187 (1984).

\bibitem{Kaplan:1983fs}
D.~B. Kaplan and H.~Georgi,
\newblock Phys. Lett. {\bf B136}, 183 (1984).

\bibitem{Georgi:2007zz}
H.~Georgi,
\newblock Comptes Rendus Physique {\bf 8}, 1029 (2007).

\bibitem{ArkaniHamed:2001nc}
N.~Arkani-Hamed, A.~G. Cohen, and H.~Georgi,
\newblock Phys. Lett. {\bf B513}, 232 (2001), hep-ph/0105239.

\bibitem{ArkaniHamed:2002pa}
N.~Arkani-Hamed, A.~G. Cohen, T.~Gregoire, and J.~G. Wacker,
\newblock JHEP {\bf 08}, 020 (2002), hep-ph/0202089.

\bibitem{ArkaniHamed:2002qy}
N.~Arkani-Hamed, A.~G. Cohen, E.~Katz, and A.~E. Nelson,
\newblock JHEP {\bf 07}, 034 (2002), hep-ph/0206021.

\bibitem{Chang:2003un}
S.~Chang and J.~G. Wacker,
\newblock Phys. Rev. {\bf D69}, 035002 (2004), hep-ph/0303001.

\bibitem{Contino:2003ve}
R.~Contino, Y.~Nomura, and A.~Pomarol,
\newblock Nucl. Phys. {\bf B671}, 148 (2003), hep-ph/0306259.

\bibitem{Agashe:2004rs}
K.~Agashe, R.~Contino, and A.~Pomarol,
\newblock Nucl. Phys. {\bf B719}, 165 (2005), hep-ph/0412089.

\bibitem{Contino:2006qr}
R.~Contino, L.~Da~Rold, and A.~Pomarol,
\newblock Phys. Rev. {\bf D75}, 055014 (2007), hep-ph/0612048.

\bibitem{Buchmuller:1985jz}
W.~Buchmuller and D.~Wyler,
\newblock Nucl. Phys. {\bf B268}, 621 (1986).

\bibitem{Mantry:2007sj}
S.~Mantry, M.~J. Ramsey-Musolf, and M.~Trott,
\newblock (2007), arXiv:0707.3152 [hep-ph].

\bibitem{Giudice:2007fh}
G.~F. Giudice, C.~, A.~Pomarol, and R.~Rattazzi,
\newblock JHEP {\bf 06}, 045 (2007), hep-ph/0703164.

\bibitem{Grinstein:2007iv}
B.~Grinstein and M.~Trott,
\newblock (2007), arXiv:0704.1505 [hep-ph].

\bibitem{Delaunay:2007wb}
C.~Delaunay, C.~Grojean, and J.~D. Wells,
\newblock JHEP {\bf 04}, 029 (2008), 0711.2511.

\bibitem{Linde:1977mm}
A.~D. Linde,
\newblock Phys. Lett. {\bf B70}, 306 (1977).

\bibitem{Dimopoulos:1978kv}
S.~Dimopoulos and L.~Susskind,
\newblock Phys. Rev. {\bf D18}, 4500 (1978).

\bibitem{Kuzmin:1985mm}
V.~A. Kuzmin, V.~A. Rubakov, and M.~E. Shaposhnikov,
\newblock Phys. Lett. {\bf B155}, 36 (1985).

\bibitem{Cohen:1993nk}
A.~G. Cohen, D.~B. Kaplan, and A.~E. Nelson,
\newblock Ann. Rev. Nucl. Part. Sci. {\bf 43}, 27 (1993), hep-ph/9302210.

\bibitem{Quiros:1999jp}
M.~Quiros,
\newblock (1999), hep-ph/9901312.

\bibitem{Sakharov:1967dj}
A.~D. Sakharov,
\newblock Pisma Zh. Eksp. Teor. Fiz. {\bf 5}, 32 (1967).

\bibitem{tHooft:1976up}
G.~'t~Hooft,
\newblock Phys. Rev. Lett. {\bf 37}, 8 (1976).

\bibitem{tHooft:1976fv}
G.~'t~Hooft,
\newblock Phys. Rev. {\bf D14}, 3432 (1976).

\bibitem{Kobayashi:1973fv}
M.~Kobayashi and T.~Maskawa,
\newblock Prog. Theor. Phys. {\bf 49}, 652 (1973).

\bibitem{PDBook}
W.-M. {Yao} {\em et~al.},
\newblock {Journal of Physics G} {\bf 33}, 1+ (2006).

\bibitem{Spergel:2006hy}
WMAP, D.~N. Spergel {\em et~al.},
\newblock Astrophys. J. Suppl. {\bf 170}, 377 (2007), astro-ph/0603449.

\bibitem{Kajantie:1996qd}
K.~Kajantie, M.~Laine, K.~Rummukainen, and M.~E. Shaposhnikov,
\newblock Nucl. Phys. {\bf B493}, 413 (1997), hep-lat/9612006.

\bibitem{Kajantie:1996mn}
K.~Kajantie, M.~Laine, K.~Rummukainen, and M.~E. Shaposhnikov,
\newblock Phys. Rev. Lett. {\bf 77}, 2887 (1996), hep-ph/9605288.

\bibitem{Kajantie:1995kf}
K.~Kajantie, M.~Laine, K.~Rummukainen, and M.~E. Shaposhnikov,
\newblock Nucl. Phys. {\bf B466}, 189 (1996), hep-lat/9510020.

\bibitem{Csikor:1998eu}
F.~Csikor, Z.~Fodor, and J.~Heitger,
\newblock Phys. Rev. Lett. {\bf 82}, 21 (1999), hep-ph/9809291.

\bibitem{Carson:1990jm}
L.~Carson, X.~Li, L.~D. McLerran, and R.-T. Wang,
\newblock Phys. Rev. {\bf D42}, 2127 (1990).

\bibitem{Shaposhnikov:1987tw}
M.~E. Shaposhnikov,
\newblock Nucl. Phys. {\bf B287}, 757 (1987).

\bibitem{Dine:1991ck}
M.~Dine, P.~Huet, and J.~Singleton, Robert~L.,
\newblock Nucl. Phys. {\bf B375}, 625 (1992).

\bibitem{Dine:1992vs}
M.~Dine, R.~G. Leigh, P.~Huet, A.~D. Linde, and D.~A. Linde,
\newblock Phys. Lett. {\bf B283}, 319 (1992), hep-ph/9203201.

\bibitem{Gavela:1993ts}
M.~B. Gavela, P.~Hernandez, J.~Orloff, and O.~Pene,
\newblock Mod. Phys. Lett. {\bf A9}, 795 (1994), hep-ph/9312215.

\bibitem{Gavela:1994dt}
M.~B. Gavela, P.~Hernandez, J.~Orloff, O.~Pene, and C.~Quimbay,
\newblock Nucl. Phys. {\bf B430}, 382 (1994), hep-ph/9406289.

\bibitem{Huet:1994jb}
P.~Huet and E.~Sather,
\newblock Phys. Rev. {\bf D51}, 379 (1995), hep-ph/9404302.

\bibitem{Fukugita:1986hr}
M.~Fukugita and T.~Yanagida,
\newblock Phys. Lett. {\bf B174}, 45 (1986).

\bibitem{Carena:1996wj}
M.~S. Carena, M.~Quiros, and C.~E.~M. Wagner,
\newblock Phys. Lett. {\bf B380}, 81 (1996), hep-ph/9603420.

\bibitem{KlapdorKleingrothaus:2001ke}
H.~V. Klapdor-Kleingrothaus, A.~Dietz, H.~L. Harney, and I.~V. Krivosheina,
\newblock Mod. Phys. Lett. {\bf A16}, 2409 (2001), hep-ph/0201231.

\bibitem{Quiros:2000wk}
M.~Quiros,
\newblock Nucl. Phys. Proc. Suppl. {\bf 101}, 401 (2001), hep-ph/0101230.

\bibitem{Mantry:2007ar}
S.~Mantry, M.~Trott, and M.~B. Wise,
\newblock (2007), arXiv:0709.1505 [hep-ph].

\bibitem{Randall:2007as}
L.~Randall,
\newblock (2007), arXiv:0711.4360 [hep-ph].

\bibitem{Noble:2007kk}
A.~Noble and M.~Perelstein,
\newblock (2007), 0711.3018.

\bibitem{Manohar:2006gz}
A.~V. Manohar and M.~B. Wise,
\newblock Phys. Lett. {\bf B636}, 107 (2006), hep-ph/0601212.

\bibitem{O'Connell:2006wi}
D.~O'Connell, M.~J. Ramsey-Musolf, and M.~B. Wise,
\newblock Phys. Rev. {\bf D75}, 037701 (2007), hep-ph/0611014.

\bibitem{Fan:2008jk}
J.~Fan, W.~D. Goldberger, A.~Ross, and W.~Skiba,
\newblock (2008), 0803.2040.

\bibitem{Goldberger:2007zk}
W.~D. Goldberger, B.~Grinstein, and W.~Skiba,
\newblock (2007), 0708.1463.

\bibitem{Profumo:2007wc}
S.~Profumo, M.~J. Ramsey-Musolf, and G.~Shaughnessy,
\newblock JHEP {\bf 08}, 010 (2007), 0705.2425.

\bibitem{Pospelov:2005pr}
M.~Pospelov and A.~Ritz,
\newblock Annals Phys. {\bf 318}, 119 (2005), hep-ph/0504231.

\bibitem{Huber:2006ri}
S.~J. Huber, M.~Pospelov, and A.~Ritz,
\newblock Phys. Rev. {\bf D75}, 036006 (2007), hep-ph/0610003.

\bibitem{Dolan:1973qd}
L.~Dolan and R.~Jackiw,
\newblock Phys. Rev. {\bf D9}, 3320 (1974).

\bibitem{Sher:1988mj}
M.~Sher,
\newblock Phys. Rept. {\bf 179}, 273 (1989).

\bibitem{Susskind:1978ms}
L.~Susskind,
\newblock Phys. Rev. {\bf D20}, 2619 (1979).

\bibitem{Sikivie:1980hm}
P.~Sikivie, L.~Susskind, M.~B. Voloshin, and V.~I. Zakharov,
\newblock Nucl. Phys. {\bf B173}, 189 (1980).

\bibitem{Barbieri:2004qk}
R.~Barbieri, A.~Pomarol, R.~Rattazzi, and A.~Strumia,
\newblock Nucl. Phys. {\bf B703}, 127 (2004), hep-ph/0405040.

\bibitem{Georgi:1984af}
H.~Georgi and D.~B. Kaplan,
\newblock Phys. Lett. {\bf B145}, 216 (1984).

\bibitem{Arzt:1994gp}
C.~Arzt, M.~B. Einhorn, and J.~Wudka,
\newblock Nucl. Phys. {\bf B433}, 41 (1995), hep-ph/9405214.

\bibitem{Manohar:1983md}
A.~Manohar and H.~Georgi,
\newblock Nucl. Phys. {\bf B234}, 189 (1984).

\bibitem{Bodeker:2004ws}
D.~Bodeker, L.~Fromme, S.~J. Huber, and M.~Seniuch,
\newblock JHEP {\bf 02}, 026 (2005), hep-ph/0412366.

\bibitem{Grojean:2004xa}
C.~Grojean, G.~Servant, and J.~D. Wells,
\newblock Phys. Rev. {\bf D71}, 036001 (2005), hep-ph/0407019.

\bibitem{Barger:2003rs}
V.~Barger, T.~Han, P.~Langacker, B.~McElrath, and P.~Zerwas,
\newblock Phys. Rev. {\bf D67}, 115001 (2003), hep-ph/0301097.

\bibitem{Chang:2003zn}
S.~Chang,
\newblock JHEP {\bf 12}, 057 (2003), hep-ph/0306034.

\bibitem{DiazCruz:2007be}
J.~L. Diaz-Cruz,
\newblock (2007), 0711.0488.

\bibitem{Maru:2006wx}
N.~Maru and K.~Takenaga,
\newblock Phys. Rev. {\bf D74}, 015017 (2006), hep-ph/0606139.

\bibitem{Coleman:1973jx}
S.~R. Coleman and E.~Weinberg,
\newblock Phys. Rev. {\bf D7}, 1888 (1973).

\bibitem{kapusta:2006pm}
J.~I. Kapusta and C.~Gale,
\newblock Cambridge, UK: Univ. Pr. (2006) 428 p.

\bibitem{Kajantie:1995dw}
K.~Kajantie, M.~Laine, K.~Rummukainen, and M.~E. Shaposhnikov,
\newblock Nucl. Phys. {\bf B458}, 90 (1996), hep-ph/9508379.

\bibitem{Rubakov:1996vz}
V.~A. Rubakov and M.~E. Shaposhnikov,
\newblock Usp. Fiz. Nauk {\bf 166}, 493 (1996), hep-ph/9603208.

\bibitem{Anderson:1991zb}
G.~W. Anderson and L.~J. Hall,
\newblock Phys. Rev. {\bf D45}, 2685 (1992).

\bibitem{fradkin}
E.~Fradkin,
\newblock Proc. Lebedev Phys. Inst. {\bf 29}, 7 (1967).

\bibitem{Weinberg:1974hy}
S.~Weinberg,
\newblock Phys. Rev. {\bf D9}, 3357 (1974).

\bibitem{Gross:1980br}
D.~J. Gross, R.~D. Pisarski, and L.~G. Yaffe,
\newblock Rev. Mod. Phys. {\bf 53}, 43 (1981).

\bibitem{Carrington:1991hz}
M.~E. Carrington,
\newblock Phys. Rev. {\bf D45}, 2933 (1992).

\bibitem{Fendley:1987ef}
P.~Fendley,
\newblock Phys. Lett. {\bf B196}, 175 (1987).

\bibitem{Espinosa:1993yi}
J.~R. Espinosa, M.~Quiros, and F.~Zwirner,
\newblock Phys. Lett. {\bf B307}, 106 (1993), hep-ph/9303317.

\bibitem{Arnold:1992rz}
P.~Arnold and O.~Espinosa,
\newblock Phys. Rev. {\bf D47}, 3546 (1993), hep-ph/9212235.

\bibitem{Weinberg:1987vp}
E.~J. Weinberg and A.-q. Wu,
\newblock Phys. Rev. {\bf D36}, 2474 (1987).

\bibitem{Heller:1997nqa}
U.~M. Heller, F.~Karsch, and J.~Rank,
\newblock Phys. Rev. {\bf D57}, 1438 (1998), hep-lat/9710033.

\bibitem{Buchmuller:1993bq}
W.~Buchmuller, Z.~Fodor, T.~Helbig, and D.~Walliser,
\newblock Ann. Phys. {\bf 234}, 260 (1994), hep-ph/9303251.

\bibitem{Zhang:1992fs}
X.-m. Zhang,
\newblock Phys. Rev. {\bf D47}, 3065 (1993), hep-ph/9301277.

\bibitem{Espinosa:1992kf}
J.~R. Espinosa, M.~Quiros, and F.~Zwirner,
\newblock Phys. Lett. {\bf B314}, 206 (1993), hep-ph/9212248.

\bibitem{Dawson:1998py}
S.~Dawson, S.~Dittmaier, and M.~Spira,
\newblock Phys. Rev. {\bf D58}, 115012 (1998), hep-ph/9805244.

\bibitem{Arnold:1994bp}
P.~Arnold,
\newblock (1994), hep-ph/9410294.

\bibitem{Linde:1980ts}
A.~D. Linde,
\newblock Phys. Lett. {\bf B96}, 289 (1980).

\end{thebibliography}

\end{document}